\documentclass{article}
\usepackage{xcolor}
\usepackage{graphics} 
\usepackage{graphicx} 
\usepackage{natbib}
\usepackage{amsmath}
\usepackage{comment}
\usepackage{amssymb}
\usepackage{listings}

\usepackage{multirow}
\usepackage{longtable}
\usepackage{pdflscape}
\usepackage{afterpage}
\usepackage{grffile}
\usepackage[colorlinks=true, citecolor=blue, linkcolor=blue, urlcolor=blue]{hyperref}
\usepackage{verbatim}
\usepackage[T1]{fontenc}
\usepackage{authblk}
\usepackage[margin=2.5cm]{geometry}

\definecolor{codegreen}{rgb}{0,0.6,0}
\definecolor{codegray}{rgb}{0.5,0.5,0.5}
\definecolor{codepurple}{rgb}{0.58,0,0.82}
\definecolor{backcolour}{rgb}{0.95,0.95,0.92}

\lstdefinestyle{mystyle}{
    backgroundcolor=\color{backcolour},   
    commentstyle=\color{codegreen},
    keywordstyle=\color{magenta},
    numberstyle=\tiny\color{codegray},
    stringstyle=\color{codepurple},
    basicstyle=\ttfamily\footnotesize,
    breakatwhitespace=false,         
    breaklines=true,                 
    captionpos=b,                    
    keepspaces=true,                 
    numbers=left,                    
    numbersep=5pt,                  
    showspaces=false,                
    showstringspaces=false,
    showtabs=false,                  
    tabsize=2
}

\title{Astrometry}
\author[1]{Gisela N.\ Ortiz-León  }
\affil[1]{\small Instituto Nacional de Astrofísica, Óptica y Electrónica}
\affil[1]{\small Apartado Postal 51 y 216, 72000 Puebla, Mexico}
\date{}

\begin{document}

\maketitle
%\tableofcontents

\begin{abstract}
Astrometry is the oldest method used to search for extrasolar planets. This technique provides key information about a planetary system, like the true mass of  planets and its three-dimensional orbital architecture. Nowadays, the space-based mission  {\it Gaia}, on one hand, and ground-based interferometry at infrared and radio wavelengths, on the other hand, can achieve astrometric accuracies of tens of micro-arseconds. Thanks to this unprecedented level of accuracy, the number of astrometrically discovered exoplanets has started to grow. In this chapter, I review the basic principles for orbital characterization of  unseen planetary companions using astrometric data. I also review the properties of the exoplanets discovered so far with astrometric methods and comment on the contributions that this technique can make to the exoplanet field.
\end{abstract}

\section{Introduction}

Astrometry, the branch of Astronomy concerned with  the registration of accurate positions of celestial objects, is the oldest method used to search for extrasolar planets.
This discipline is as old as Astronomy itself. Many of the ancient civilizations observed and registered the positions of stars and other celestial bodies.
The greek astronomer Hipparchus compiled coordinates of celestial objects possibly in 135 BCE and his catalog is considered the earliest attempt to chart the entire sky observable with the naked eyed \citep{Toomer,Marchant2022Natur}. Stellar coordinates in Hipparchus star catalog were accurate to within $1$ degree \citep{Gysembergh2022}. Since then, the astrometric accuracy has remarkably improved, reaching levels of tens of microarseconds as in the {\it Gaia} astrometric space mission \citep{Gaia2023} or from the ground by using  interferometric techniques \citep{Reid2014,Eisenhauer2023}. This achievement represents an eight orders of magnitude increase with respect to the accuracy of the Hipparchus' star catalog. 

\begin{figure}[!th]
\begin{center}
 \includegraphics[width=0.99\textwidth,angle=0]{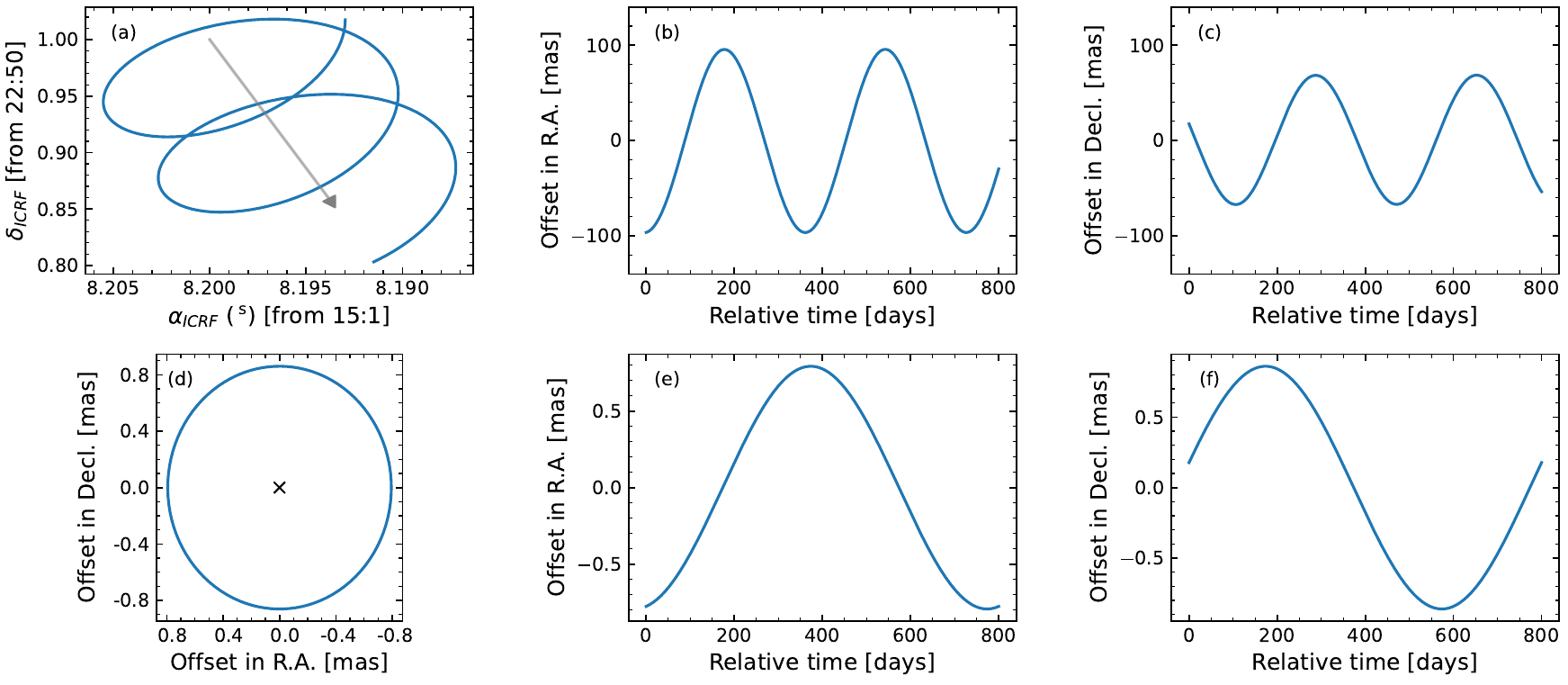}
\caption{Sky motion of a star with a planet at a distance $d=100$~pc. The star is at position $\alpha$=15$^{\rm h}$01$^{\rm m}$08.2$^{\rm s}$, $\delta$=22$^{\rm d}$50$^{\rm m}$01.0$^{\rm s}$ at epoch 2019.62 and its mass is $M_\star=0.08~{\rm M}_\odot$. The proper motion of the star-planet barycenter is ($\mu_\alpha\cos\delta$,~$\mu_\delta$) = ($-43.1$,$-65.43$)~mas~yr$^{-1}$. The planet, with a mass of $M_p=1~{\rm M}_{\rm Jup}$, is in a circular (eccentricity, $e$ = 0) and face-on orbit ($i$ = 0) around the host-star with a period ($P$) of 800 days. The semi-major axis of the planet is $a_p = [ M_\star^3 P^2/(M_\star + M_p)^2 ]^{1/3}$ = 0.72~au.
In panel a) the blue line shows the total motion of the star that results from the parallax effect and the proper motion, which is negative in both directions and indicated by the gray line. Panels b) and c) show  R.A. and Decl.\ offsets as a function of time, after removing the proper motion. The resulting signal has a period of 365.25~days, and corresponds to the effect of parallax. Panel d) shows the star sky motion after removing the contributions from parallax and proper motions. The star describes a circular orbit around the star-planet barycenter (black cross), with a semi-major axis of $M_p \cdot a_p / d \cdot M_\star$ = 0.86~mas (see Section \ref{sec:method}). Panels e) and f) show the same R.A. and Decl.\ offsets as in panel d) but as a function of time. The signals have a period of 800~days. All offsets are relative to the star position at the reference epoch.    } 

\label{fig:model}
\end{center}
\end{figure}

By registering the positions of a star at a series of times
an observer on Earth can measure the star path on the plane of the sky (Fig. \ref{fig:model}). For isolated stars, the contributions to this path are an effect of parallax and its {\it proper motion}. The parallax, $\varpi ['']$, is the apparent angular shift owing to the translation movement of the Earth, with a period of 1 year,
and is equal to $1''/d[\rm{pc}]$, where $d$ is the source distance in parsecs. 
If a star has one companion (which could be a star, a brown dwarf  or a planet), the two bodies in the system follow Keplerian orbits around their common center of mass (commonly referred to as the barycenter). 
Then, an additional contribution to the apparent path of the star is due to the orbital motion around the star-companion barycenter (see panel d, Fig. \ref{fig:model}). Thus, in a planetary system the planets perturb the sky motion of the host star. The amplitude of this perturbation is known as the astrometric signature, $\alpha_\star$, and corresponds to the apparent semi-major axis of the star orbit. In other words, the astrometric signature is the wobble of the host star due to the presence of a companion. This signature, which is proportional to the companion mass, can be quite small ($<1$ milli-arcsecond) for planetary-mass companions, even for the nearest stars.

Precise measurements of the path on the plane of the sky are used for the determination of the parallax, and provide a direct measurement of the distance to a star. 
The measurements of the distance to stars using the parallax method by Bessel, von Struve, and Henderson in the 1830s marked the beginning of modern astrometry \citep{Bessel1838AN,Bessel1838MNRAS,vonStruve,Reid2020}.

While astrometry was a successful technique for characterizing astrometric binary star systems, which dates back to late 18th century, its use for detecting companions with masses below 0.1 $M_\odot$ was cast into doubt by many astronomers. Statements on the astrometric discovery of an unseen body in the binary system 70 Ophiuchi were made by  William Stephen Jacob in 1855 \citep{Jacob1855}. Although this claim was shown to be erroneous \citep{Heintz1988}, it marked the earliest known attempt to detect an extrasolar planet. 
Efforts to hunt exoplanets using astrometric methods were made by others for another century, leading to controversial claims \citep[e.g.][]{Strand1943,Lippincott1960,vandeKamp1969}. However, all these discoveries were subsequently refuted, testifying the significant difficulties involved in astrometric exoplanet searches with ground-based optical telescopes. 

Measuring highly reliable parallaxes and sky motions from the ground is challenging. Earth’s atmosphere introduces distortion effects that are difficult to overcome, imposing a severe limit on the astrometric precision achievable with  ground-based instruments. 
It was not until the launch of the  European Space Agency's  \textsc{Hipparcos} mission, in 1989, that accurate astrometric data for a significant number of stars were obtained. \textsc{Hipparcos} achieved an astrometric precision of 1 milli-arcsecond \citep{Eisenhauer2023} and parallaxes were determined to better than 10\% for about 20,000 objects. 
Although \textsc{Hipparcos} made no exoplanet discoveries through the detection of the star astrometric signature due to planetary-mass companions, it contributed to the characterization of planetary systems known at the time (see \citealt{Perryman2008} for a review on the \textsc{Hipparcos} science). 
For example, \cite{Mazeh1999} combined orbital elements derived by the radial velocity (RV) method {(see Chapter 2)} with astrometric data provided by \textsc{Hipparcos} to constrain the motion of the barycenter, the astrometric signature, and a subset of the orbital parameters of the outer companion in the triple planetary system $\upsilon$~And.  
The mass of this outer companion and estimates of the mass of the two inner planets were also obtained. 

It is thanks to the development of cutting-edge instrumentation, placed on board of space-based observatories, and  the progress on advanced techniques at  wavelengths other than optical that  astrometry has reemerged as a technique that can uniquely contribute to the field of exoplanet exploration. A few exoplanets have been already  detected with astrometric methods and the number will continue to grow as the astrometric precision  improves.

This chapter describes the analysis methods used to detect and characterize unseen planetary companions from astrometric data. The process of fitting a Keplerian orbit is presented for the general case of a single planet orbiting around a single star. 
Particular cases, for instance fitting  a planetary orbit in a binary star system, are also discussed to a lesser extent.  I also describe the differences between absolute and relative astrometry, and provide a short overview of the observation principles of {\it Gaia}  and of radio interferometers. Discoveries made with the proper motion anomaly and infrared astrometry are also discussed. 
Then, I review the demographics of planetary-mass objects discovered with astrometric methods and how a substantial sample of astrometrically detected planets can provide valuable insights into planet formation and evolution. Finally, an example of data analysis is presented in the tutorial, where synthetic astrometric data are used to exemplify the use of radio interferometry as a tool for characterization of planetary orbits. 

The reader is invited to see \cite{Brandt2024} for a comprehensive review on the {\it Gaia} astrometric capabilities for exoplanet discovering.  A summary of the historical contributions of astrometry to the study of exoplanets can be found in \cite{Malbet2018}. A comprehensive discussion of other detection methods is presented in the remaining chapters of this book. 

\section{Method description}
\label{sec:method}

The astrometric signature, $\alpha_\star$, is the amplitude of the reflex motion of a planet-host star induced by the gravitational influence of a planet with mass $M_{\rm p}$.  After removing the contribution of parallax and proper motion, both the star and the planet describe elliptical paths on the plane of the sky, that result from their orbital movement around the star-planet barycenter. The angular size in arcseconds of the semi-major axis of the star orbit is given by, 

\begin{equation}\label{eq:asignature}
\alpha_\star = \frac{M_{\rm p}}{M_\star} \cdot \frac{a_{\rm p}}{d} 
\end{equation}

\begin{figure}[!th]
\begin{center}
 \includegraphics[width=0.7\textwidth,angle=0]{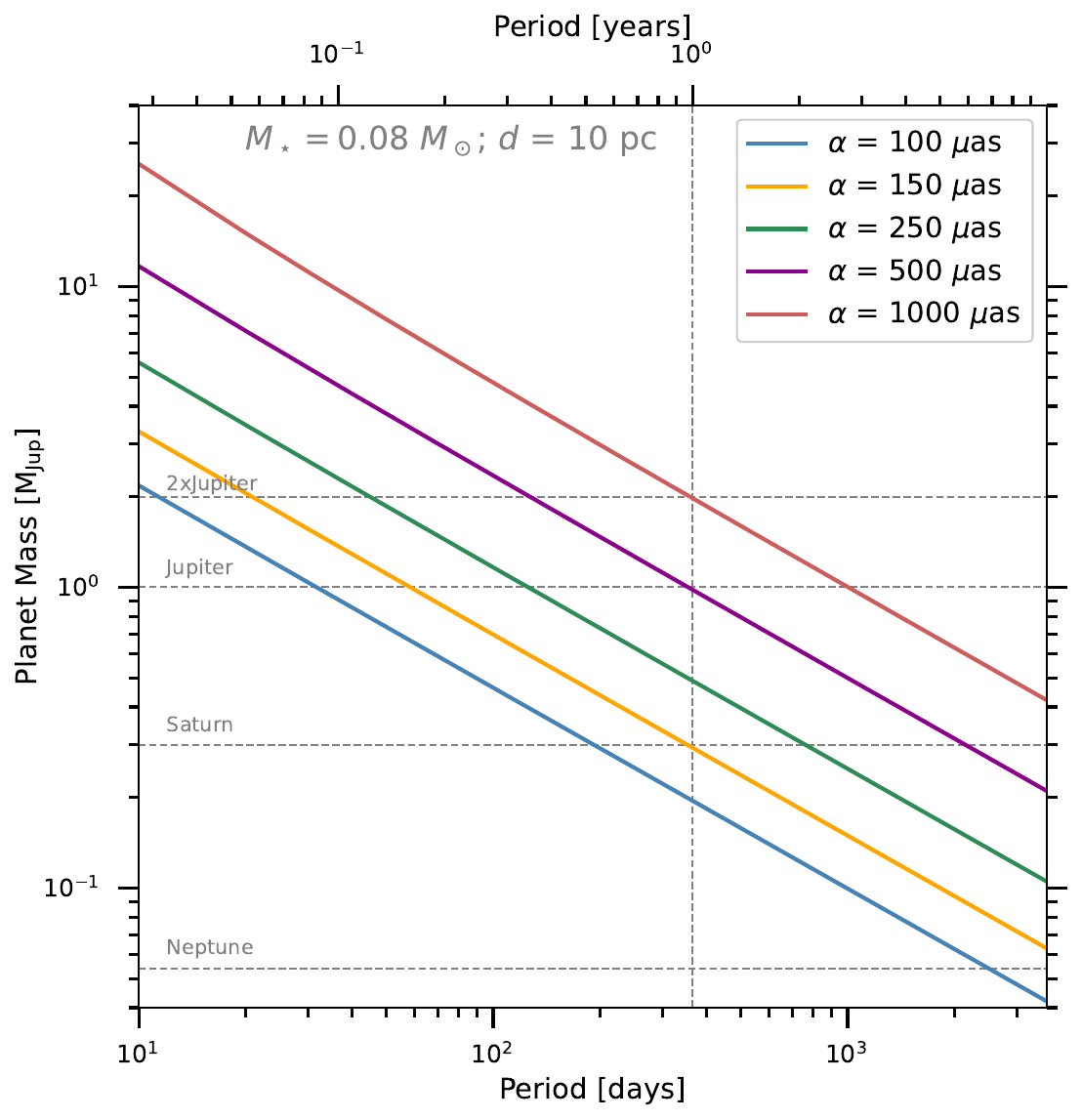}
\caption{Relation between planet mass and orbital period for values of the astrometric signature as indicated in the legend. The host star has a mass of $M_\star=0.08~M_\odot$ and is at a distance $d=10$~pc. The horizontal lines indicate companions of different masses. The vertical line correspond to a period of 1~yr. 
}
\label{fig:signature}
\end{center}
\end{figure}

\noindent where $M_\star$ is the mass of the host star, $a_{\rm p}$ is the semi-major axis of the planet's orbit in AU and $d$ the distance in parsecs. Exoplanet searches with the astrometry method aim to disentangle this signature from the observed total motion of the star, which, at nearby distances ($\sim$10~pc), is largely dominated by parallax and proper motion.

Equation (\ref{eq:asignature}) implies that, for a fixed star mass, the astrometric signature is more prominent for massive planets, for planets with large orbital periods, and for nearby stars. To investigate the space of parameters probed by a given limit of the astrometric precision, we combine this equation with Kepler's Third Law,

\begin{equation}
M_{\rm p} = \frac{4 \pi^2}{G} \frac{d^3 (\alpha_\star + \alpha_{\rm p})^2 \alpha_{\star}}{P^2},
\end{equation}

\noindent resulting in the following expression,

\begin{equation} \label{eq:Mp}
M_{\rm p}^3 = \frac{4 \pi^2}{G} \frac{d^3}{P^2} \alpha_\star^3 (M_\star + M_{\rm p})^2, 
\end{equation}

\noindent where $\alpha_{\rm p} = a_{\rm p} / d $ and $G$ is the Gravitational constant. This expression can be solved for the planet mass by fixing the value of the astrometric signature. In Fig.~\ref{fig:signature}, which considers the case of a single companion orbiting around an isolated star, the star mass is fixed to 0.08~M$_\odot$ and the distance to 10~pc. The solid lines show the planet mass as a function of the orbital period\footnote{It would be useful to clarify that what is actually measured is the orbital period of the star around the star-planet barycenter. However, since this is identical to the planet's orbital period, it is simply referred to as the orbital period.}, after solving Eq.~(\ref{eq:Mp}) with $\alpha_\star =$~100, 200, 500 and 1000~micro-arcseconds ($\mu$as).

Looking at the uppermost horizontal line in the plot we see that, for a planet with a mass of $M_p \geq 2$~M$_{\rm Jup}$, only periods $P\geq1$~yr would induce an astrometric signature above 1~mas, while Jupiter- and Saturn-mass planets in a $1$-yr orbit produce signatures of 150 -- 500~$\mu$as around a star with 0.08~M$_\odot$.  Since the detection of a planet requieres a  signal-to-noise ratio of 3--5  in the star's reflex motion, it is clear that the detection of such signatures requires an astrometric precision at the sub-milliarsecond level. Indeed, astrometric errors of $\sim$ 50 micro-arcseconds are needed for  the unambiguous detection of Jupiter and Saturn-like planets around  stars with masses $\sim 0.08~M_\odot$.  

The characterization of exoplanetary orbits with astrometry is similar to the orbital characterization of binary systems, although there are some subtle differences. First, in binary systems usually both stars are visible. As is discussed in section \ref{sec:rel-astrometry}, the positions of the secondary star with respect to the primary trace the {\it relative} orbit. Fitting the relative orbit yield the orbital elements, and the total mass of the system if the distance is known. The semi-major axis of the relative orbit, $a$, satisfies  $a=a_1+a_2$, where $a_1$ and $a_2$ are the semi-major axis of the primary and secondary's orbit around the system's barycenter, respectively.

In the case of a star-planet system usually only the star is visible, thus we  characterize the star's orbit around the star-planet barycenter. If the star positions are measured in an {\it absolute} reference frame (i.e., not relative to a nearby star), the contributions  to the sky path of the star  by the parallax and proper motion should be taken into account. From astrometric data of the star alone it is possible to directly derive $a_1$ and the rest of the Keplerian elements such as orbital period, inclination and eccentricity (see Section \ref{sec:absolute-astrometry}). The mass and semi-major axis of the planet can be obtained through Kepler's third law if  a measurement of the mass of the host star is available. It is important to note that the mass of the stellar host is usually estimated from isochrone fitting, so they are model-dependent. Only in binary systems can the mass be measured dynamically.  

It is convenient to introduce now the nomenclature used for orbital characterization. I first consider the case of a visual binary in Section \ref{sec:rel-astrometry}. This will help the reader to recognize the elements of a Keplerian orbit. The same nomenclature will be used in Section \ref{sec:absolute-astrometry}, where I describe the methodology behind the characterization of exoplanetary orbits.

\subsection{Relative astrometry}\label{sec:rel-astrometry}

For visual binaries, the position of the secondary is measured relative to the primary star. In polar coordinates, the position is given by the angular separation, $\rho$, and the position angle, $\theta$, measured from north in an anticlockwise
direction as seen by the observer. The separation and position angle will change over time, tracing the orbital motion of the binary projected on the sky. If the orbital motion is in the anticlockwise direction, the orbit is said to be prograde. If the orbital motion is in clockwise direction, the orbit is said to be retrograde. 

\begin{figure}[!th]
\begin{center}
 \includegraphics[width=0.7\textwidth,angle=0]{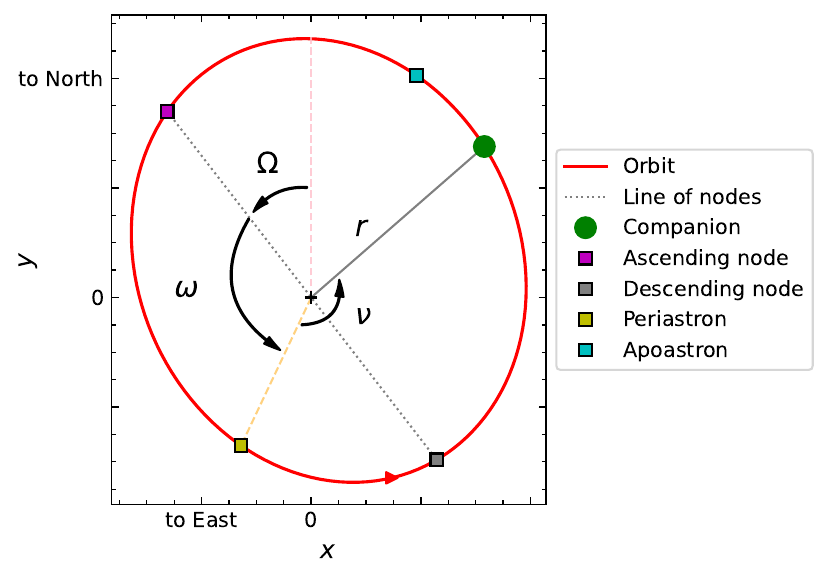}
\caption{Representation of a Keplerian orbit projected on the plane of the sky. In the case of a visual binary, it is the relative orbit of the secondary around the primary. In the case of a star-planet system, it is the orbit of the star around the star-planet barycenter. The position angle of the ascending node ($\Omega$) is measured from North (the pink dashed line is drawn to guide the eye) through East.  The argument of periastron ($\omega$) is the angle between the ascending node and the periastron (whose direction is indicated by the yellow dashed line) measured in the direction of motion. The true anomaly ($\nu$) is the angle between the instantaneous
companion’s location and the direction of the periastron. 
The $x$-axis points to East and the $y$-axis points to North. The direction of orbital motion is indicated by the red arrow. 
The  ``+''  marks the position of the primary star (in the case of a visual binary) or the host star (in the case of a star-planet system). 
}
\label{fig:sky-orbit}
\end{center}
\end{figure}

\begin{figure}[!th]
\begin{center}
 \includegraphics[width=0.9\textwidth,angle=0]{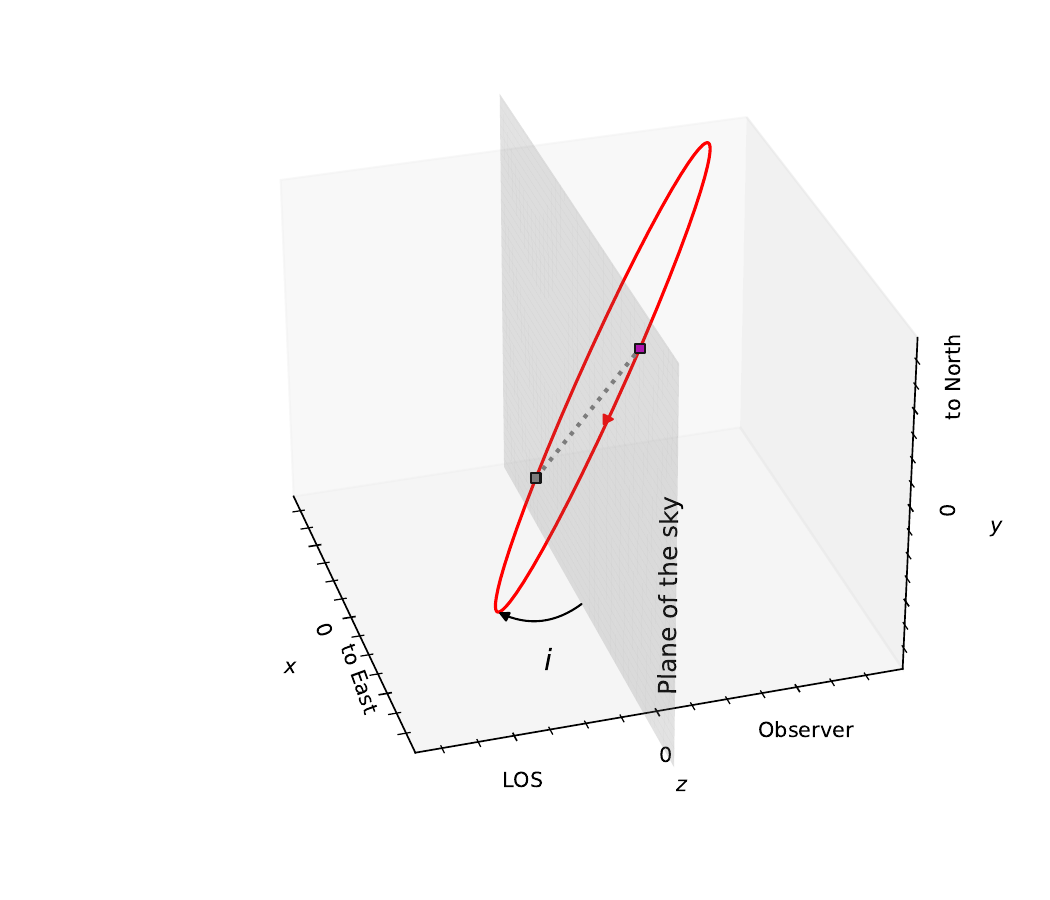}
\caption{Three-dimensional representation of a Keplerian orbit. The magenta and grey squares mark the ascending and descending node, respectively. The dotted line is the line of nodes. The observer is located on the negative $z$ axis. The angle $i$ denotes the inclination of the orbit. }
\label{fig:3D-orbit}
\end{center}
\end{figure}

To introduce the elements of a Keplerian orbit, 
we adopt a coordinate system as in \cite{Green1985} and  illustrated in Fig.~\ref{fig:sky-orbit}. 
The $+x$ direction points East and the $+y$ direction points North. The $z$-axis is perpendicular to the $x-y$ plane, this is, to the sky plane, forming a right-handed system. Therefore, the line of sight (LOS) points in the $+z$ direction, while the  $-z$ direction points to the observer. The reference plane is the sky plane. Figure \ref{fig:3D-orbit} illustrates an example of a Keplerian orbit in the three-dimensional space (the true orbit), with the direction of companion motion indicated by the red arrow. The projection of this orbit on the reference plane is called the apparent orbit and is illustrated in Fig.~\ref{fig:sky-orbit}.  The true orbit intersects the reference plane in two nodes. The ascending node is the point where the companion ``crosses'' the plane of the sky away from the observer. The position angle of the ascending node  (measured from north via east) is denoted as $\Omega$ (Fig.~\ref{fig:sky-orbit}). The line joining the two nodes is the line of nodes. The inclination angle, $i$, is the angle between the orbital plane and the reference plane. An orbit with zero inclination is located in the $x-y$ plane. If the inclination increases, the orbit rotates around the $y$-axis, so that $+x$ is rotated toward $+z$ (Fig.~\ref{fig:3D-orbit}). The epoch of periastron passage, $\tau$, is the time corresponding to periastron, when the companion is closest to the primary. The argument of periastron, $\omega$, is the angle between the ascending node and the periastron measured in the direction of motion. 

In this coordinate system the coordinates of the companion with respect to the primary are given by \citep{Green1985},
\begin{equation}\label{eq:rel_x}
x = r [\cos(\nu+\omega)\sin{\Omega} + \sin(\nu+\omega)\cos{\Omega}\cos{i}],
\end{equation}
\begin{equation}\label{eq:rel_y}
y = r [\cos(\nu+\omega)\cos{\Omega} - \sin(\nu+\omega)\sin{\Omega}\cos{i}],
\end{equation}
\begin{equation}\label{eq:rel_z}
z = r \sin(\nu+\omega)\sin{i},
\end{equation}

\noindent where $r$ is the radius vector of the instantaneous companion's location and $\nu$ is the true anomaly, which corresponds to the angle between the instantaneous companion's location  and the direction of the periastron (Fig.~\ref{fig:sky-orbit}). 

Measurements of $\rho$ and $\theta$ can be converted to $(x,y)$ positions according to,
\begin{equation}
 x = \rho\sin{\theta},
\end{equation}
\begin{equation}
 y = \rho\cos{\theta}.
\end{equation}

The series of values $(x,y,t)$ are then used to determine the apparent orbit, whose analytical expression is given by equations (4)--(5). However, notice that these equations are not linear in all the orbital elements. The Thiele-Innes method \citep{Thiele1883,Binnendijk1960,Heintz1978} is a formulation to transform equations (4)--(5) such that they become linear in a subset of the  orbital parameters. In this formulation (4) and (5) can be rewritten as \citep{Green1985},

\begin{equation}\label{eq:TI_ra}
x = X1 \cdot\left( \frac{r}{a}\right)\cos \nu + X2\cdot \left( \frac{r}{a}\right)\sin \nu,
\end{equation}
\begin{equation}\label{eq:TI_dec}
y = Y1\cdot \left( \frac{r}{a}\right)\cos \nu  + Y2\cdot \left( \frac{r}{a}\right)\sin \nu,
\end{equation}

\noindent where $X1$, $X2$, $Y1$ and $Y2$ are known as the Thiele-Innes constants, which are given by,

\begin{equation}\label{eq:X1}
X1 = a \cdot( \cos\omega \sin\Omega + \sin\omega\cos\Omega\cos i),
\end{equation}
\begin{equation}\label{eq:Y1}
Y1 = a \cdot ( \cos\omega \cos\Omega - \sin\omega\sin\Omega\cos i),
\end{equation}
\begin{equation}\label{eq:X2}
X2 = a \cdot ( -\sin\omega \sin\Omega + \cos\omega\cos\Omega\cos i),
\end{equation}
\begin{equation}\label{eq:Y2}
Y2 = a \cdot ( -\sin\omega \cos\Omega - \cos\omega\sin\Omega\cos i).
\end{equation}

The radius vector of the orbit,  $r$, the true anomaly, $\nu$, and the eccentricity, $e$, are related through the equations,

\begin{equation}\label{eq:dyneq}
\frac{r}{a} = \frac{1-e^2}{1+e \cos\nu} = 1 - e\cos E, 
\end{equation}
\bigskip

\noindent where $a$ is the semi-major axis and $E$ is the eccentric anomaly. This angle corresponds to the angle between the direction of the periastron and the projection of the companion’s location onto a circle of radius $a$ that intersects the orbit at periastron.  The eccentric anomaly is the solution to the following expression, known as  Kepler's equation

\begin{equation}\label{eq:Kepler}
E - e \sin E = \frac{2\pi(t-\tau)}{P},
\end{equation}

\noindent where $\tau$ is the epoch of periastron passage and $P$ the orbital period, while the true anomaly $\nu$ satisfies the condition 

\begin{equation}\label{eq:nu}
\tan \left( \frac{\nu}{2} \right) = \left( \frac{1+e}{1-e}\right)^{1/2} \tan \left( \frac{E}{2} \right). 
\end{equation}

For a given set of parameters $e$, $P$, and $\tau$, the true anomaly can be obtained for any point in the orbit by solving equation (\ref{eq:Kepler}) for $E$ and substituting it into equation (\ref{eq:nu}). Then $r/a$ is found using equation (\ref{eq:dyneq}), which can be substituted into equations (\ref{eq:TI_ra})--(\ref{eq:TI_dec}).

\subsection{Absolute astrometry}\label{sec:absolute-astrometry}

I will now introduce the fundamental equations to model astrometric data 
of the sort that are provided by Radio Interferometry or {\it Gaia}, noting that the approach followed to fit a planetary orbit is also valid for stellar companions and brown dwarfs.
We assume that positions are measured relative to an {\it absolute} reference frame, this is, a well defined celestial coordinate system, and not relative to nearby stars.

The equations that describe the sky path of a isolated star hosting a single planet are

\begin{equation}\label{eq:ra_total}
\alpha(t) = \alpha_0 + \mu_\alpha \cos\delta \cdot (t - t_0) + \varpi \cdot f_\alpha(t) + G_\alpha(t), 
\end{equation}
\begin{equation}\label{eq:dec_total}
\delta(t) = \delta_0 + \mu_\delta \cdot (t - t_0) + \varpi \cdot f_\delta(t) + G_\delta(t),
\end{equation}

\noindent where $(\alpha_0,\delta_0)$ is a reference position at a reference time $t_0$, $(\mu_\alpha,\mu_\delta)$ are the proper motion components of the star-planet barycenter, and $\varpi$ is the parallax. The terms $f_\alpha$ and $f_\delta$ are the projections of the parallactic ellipse over right ascension and declination, respectively, and are given by \citep{Seidelmann1992},

\begin{equation}\label{eq:earth_ra}
f_\alpha(t) = (X_{\rm E} \sin \alpha_1 - Y_{\rm E} \cos\alpha_1 )/15 \cos\delta_1,
\end{equation}
\begin{equation}\label{eq:earth_dec}
f_\delta(t) = X_{\rm E} \cos \alpha_1  \sin \delta_1 + Y_{\rm E} \sin\alpha_1 \sin \delta_1 - Z_{\rm E} \cos \delta_1,
\end{equation}

\noindent where $(X_{\rm E},Y_{\rm E},Z_{\rm E})$ are the barycentric coordinates of the Earth in AU, and where $\alpha_1=\alpha-\pi f_\alpha(t)$ and $\delta_1=\delta-\pi f_\alpha(t)$ are the coordinates of the barycentric place of the star at each epoch \citep[e.g.,][]{Loinard2007}.  
The terms $G_\alpha(t)$ and $G_\delta(t)$ are the components of the Keplerian motion around the star-planet barycenter, which can be expressed
in terms of the Thiele--Innes constants $X1$, $Y1$, $X2$, and $Y2$,

\begin{equation}\label{eq:G_ra}
G_\alpha(t) = X1 \cdot\left( \frac{r}{a_1}\right)\cos \nu + X2\cdot \left( \frac{r}{a_1}\right)\sin \nu,
\end{equation}
\begin{equation}\label{eq:G_dec}
G_\delta(t) = Y1\cdot \left( \frac{r}{a_1}\right)\cos \nu  + Y2\cdot \left( \frac{r}{a_1}\right)\sin \nu.
\end{equation}

\noindent where $a_1$ is the semi-major axis of the star orbit around the star-planet barycenter. 
Inserting equations (\ref{eq:G_ra}) and (\ref{eq:G_dec}) into (\ref{eq:ra_total}) and (\ref{eq:dec_total}) yields

\begin{equation}\label{eq:ra_total2}
\alpha(t) = \alpha_0 + \mu_\alpha \cos\delta \cdot(t - t_0) + \varpi \cdot f_\alpha(t) + X1\cdot \left( \frac{r}{a_1}\right)\cos \nu + X2\cdot \left( \frac{r}{a_1}\right)\sin \nu,
\end{equation}
\begin{equation}\label{eq:dec_total2}
\delta(t) = \delta_0 + \mu_\delta \cdot (t - t_0) + \varpi \cdot f_\delta(t) + Y1 \cdot \left( \frac{r}{a_1}\right)\cos \nu  + Y2 \cdot \left( \frac{r}{a_1}\right)\sin \nu.
\end{equation}

Given a set of measurements of the star position $(\alpha_1,\delta_1)$, $(\alpha_2,\delta_2)$, ..., $(\alpha_N,\delta_N)$ at epochs $t_1$,  $t_2$, ... , $t_N$, equations (\ref{eq:ra_total2}) and (\ref{eq:dec_total2}) can be used to find the best astrometric parameters ($\alpha_0$, $\delta_0$, $\mu_\alpha$, $\mu_\delta$, and $\varpi$) and orbital elements (contained in the Thiele--Innes constants) that fit the data. Again, notice that (\ref{eq:ra_total2}) and (\ref{eq:dec_total2}) are not linear in all the orbital elements. In order to linearize these equations in terms of the five astrometric parameters and the four Thiele--Innes constants one can fix the triplet ($e$, $P$, $\tau$), find $E$ by solving equation (\ref{eq:Kepler}) and then find $\nu$ using (\ref{eq:nu}) and $r/a$ using (\ref{eq:dyneq}). 

The free parameters in Eqs.\ (\ref{eq:ra_total2}) and (\ref{eq:dec_total2}) are $\alpha_0$, $\delta_0$, $\mu_\alpha$, $\mu_\delta$, $\varpi$, $X1$, $Y1$, $X2$, $Y2$, which may be found by minimizing $\chi^2$. The orbital elements $a_1$, $\omega$, $\Omega$, and $i$ are then derived from the Thiele-Innes constants, by combining equations (\ref{eq:X1})--(\ref{eq:Y2}); see \cite{Green1985}. To fit $e$, $P$, and $\tau$ one can implement a Monte-Carlo approach as in \cite{Curiel2019,Curiel2020}, where the smallest $\chi^2$ is searched on finite grids of $e$, $P$, and $\tau$.

It should be noted that adding $180^{\rm o}$ to both $\omega$ and $\Omega$ will not alter the right sides of equations (\ref{eq:X1})--(\ref{eq:Y2}). This ambiguity of $180^{\rm o}$ arises from the fact that  astrometric data do not provide information of the star location along the $z$-axis. Therefore, these angles cannot be unambiguously determined from astrometric data alone. Radial velocity data are needed to solve the ambiguity.

\subsubsection{Radio Interferometry}

There are few examples of orbital fitting by means of absolute astrometric measurements. TVLM~513--46546b \citep{Curiel2020} and  GJ~896Ab \citep{Curiel2022} were found from astrometric data taken with the Very Long Baseline Array (VLBA) at radio wavelengths. The VLBA is an interferometric array consisting of 10 25-m identical antennas spread across the United States. The data from the individual antennas are combined to form a virtual telescope of $\sim8,600$~km in diameter, reaching an angular resolution of about 1~mas at an observing wavelength of 5~cm. 

Astrometric observations with Very Long Baseline Interferometry (VLBI) consists of measuring the angular offset between the observed source and one or more nearby calibrators, typically within a few degrees from the target \citep{Reid2022}. These observations provide absolute stellar positions because  calibrator sources are usually distant quasars with negligible proper motion. 

The fundamental quantity of a radio interferometer is the geometric delay determined by the arrival time difference at the the locations of the antennas,
\begin{equation}
    \tau_g = \frac{\vec s\cdot \vec B}{c},
\end{equation}

\noindent where $\vec B$ is the separation vector between two antennas, $\vec s$ the vector pointing towards the direction of the observed source and $c$ is the speed of light. Given an uncertainty in the delay measurement $\Delta\tau$, the uncertainty on the sky position of a star (astrometric error) can be estimated as \citep{Reid2014},

\begin{equation}\label{eq:astro-radio}
  \Delta s =  \theta_{\rm sep} \frac{c\Delta\tau}{|B|}
\end{equation} 

\noindent where $\theta_{\rm sep}$ is the angular offset between the target and the calibrator. 
For a typical VLBI observation at $\sim 8$~GHz, $|B|\sim$~8,000~km, $\theta_{\rm sep}\sim$~2~deg and $c\Delta\tau\sim5$~cm, resulting in astrometric errors of $\Delta s\sim50~\mu{\rm as}$.

Equation \ref{eq:astro-radio} ignores the effects caused by stable contributions and short-term fluctuations of the total electron content in the Earth's ionosphere, which are important at low radio frequencies ($\nu<8$~GHz) and propagate into a position error in the source. The astrometric errors arising from these effects scale as $\nu^{-2}$, resulting in values $\gtrsim 1$~mas at 1.4~GHz \citep{Rioja2020}. 
Advanced calibration strategies have been developed to mitigate atmospheric effects --the dominant source of astrometric error-- and thereby improve astrometric accuracy at both high and low radio frequencies. For a comprehensive review on astrometric calibration methods we refer the reader to \cite{Rioja2020}. 
Here, it suffices to mention that several of these strategies have already been implemented in VLBI observations, leading to significant improvement in the astrometric accuracy (e.g., achieving $\sim100~\mu$as at 1.6~GHz; \citealt{Brisken2002}), while next generation of methods will enable ultra-precise astrometry (at the $\mu$as level) across a wide range of frequencies when implemented in next-generation telescopes. Chapter {10} will provide a more detailed discussion of the future prospects for ultra-precise astrometry with forthcoming facilities.

\subsubsection{ESA {\it Gaia} mission } 

Before naming exoplanets astrometrically detected by {\it Gaia}, it is important to mention a subtle difference when modeling {\it Gaia} time series as opposed to modeling astrometric data from very long baseline interferometry,
which is due to the nature of {\it Gaia} measurements.

The basic operation principle of {\it Gaia} follows its predecessor mission, \textsc{Hipparcos} \citep{Perryman_2008}. The {\it Gaia} satellite is equipped with two fields of view, separated by a constant and large angle on the sky, which is essential for the derivation of absolute trigonometric parallaxes. By slowly rotating about a spin axis perpendicular to the scanning circle, 
{\it Gaia} measures the crossing times of targets transiting the instrument focal plane. 
The crossing times are associated to one-dimensional, ``along-scan'' positions of the target photocentre, denoted as $w$. These measurements are taken in a 2D tangential coordinate system, where the origin is at a reference equatorial position, and the axis of the longitude coordinate is oriented along the scanning direction \citep{Gaia2016,GC2024}. The motion of the star in right ascension and declination is projected onto the scan direction. 
Thus, for a single source the astrometric model is written as \citep{Holl2023},

\begin{equation}\label{eq:gaia-ss}
 w_{ss} = (\Delta\alpha^\star  + \mu_{\alpha^\star} t)\sin\psi + (\Delta\delta  + \mu_{\delta} t)\cos\psi + \varpi f_\varpi,
\end{equation}

\noindent where $w_{ss}$ are the along-scan measurements at times $t$, $\Delta\alpha^\star = \Delta\alpha\cos\delta$ and $\Delta\delta$ are offsets from a reference point along right ascension and declination, $\mu_{\alpha^\star} = \mu_{\alpha} \cos\delta$ and $\mu_{\alpha}$ are the proper motion components in these directions, $\varpi$ is the parallax and $\psi$ is the scan angle, which is the angle of the scanning direction with respect to local North. The term $f_\varpi$ is the projection factor of the parallactic ellipse over the scanning direction. %this direction. 

For a star hosting a single planet, the term due to the Keplerian motion around the star-planet barycenter should be added to equation (\ref{eq:gaia-ss}). Then, the equation that fully describes the star  motion is \citep{Holl2023},

\begin{equation}\label{eq:gaia-full}
\begin{split}
 w = (\Delta\alpha^\star  + \mu_{\alpha^\star}~t)\sin\psi + (\Delta\delta  + \mu_{\delta}~t)\cos\psi +  \varpi f_\varpi + \\
 (X1~X + X2~Y)\sin\psi + (Y1~X + Y2~Y)\cos\psi,
 \end{split}
\end{equation}

\noindent where $X1$, $X2$, $Y1$ and $Y2$ are the Thiele--Innes constants and,

\begin{equation}
 X = \cos E - e
\end{equation}
\begin{equation}
 Y = \sqrt{1-e^2} \sin E.
\end{equation}

As stated in section \ref{sec:rel-astrometry}, the eccentric anomaly $E$ satisfies the Kepler's equation given by (\ref{eq:Kepler}). Thus, the model given by (\ref{eq:gaia-full}) has 12 free parameters: 5 astrometric parameters from the single-source model ($\Delta\alpha^\star$, $\Delta\delta$, $ \mu_{\alpha^\star}$, $\mu_{\delta}$, and $\varpi$), and 7 parameters from the Keplerian orbit ($X1$, $X2$, $Y1$, $Y2$, $e$, $P$ and $\tau$). A detailed description of the {\it Gaia} processing pipelines for fitting these parameters 
is provided in \cite{Halbwachs2023} and \cite{Holl2023}. In essence, the linear parameters $\Delta\alpha^\star$, $\Delta\delta$, $ \mu_{\alpha^\star}$, $\mu_{\delta}$, $\varpi$ $X1$, $X2$, $Y1$, and $Y2$ are fitted directly by means of eq.~(\ref{eq:gaia-full}), while the three non-linear parameters $e$, $P$ and $\tau$ are found using different algorithms. 

{\it Gaia} full time series are expected to be released by the  Collaboration in Data Release 4. An example of {\it Gaia} measurements can be found in \cite{GC2024}, where a black hole in a binary system was identified by fitting the orbit of the visible star in the system. The authors published the data used to produce the orbital solution, which consist of barycentric time $t$, along-scan positions of the photocentre $w$, scan angle $\psi$, parallax factor $f_w$, as well as the associated uncertainties $\Delta w$.

\bigskip

Along with the publication of the Data Release 3 (DR3), the {\it Gaia} Collaboration released the first catalogue of non-single star solutions. These are stars whose positions cannot be well fitted with the five astrometric parameters model given by Eq.\ (\ref{eq:gaia-ss}). The  catalogue includes orbital solutions of 72 astrometric companions in the  mass range ($M_p<20$~$M_{\rm Jup}$), of which 9 are known exoplanets \citep{GC2023AA674A34G}. 
Among these, Gaia-4b was successfully confirmed with follow-up RV observations \citep{Stefansson2025}.  With an estimated mass of $11.8\pm0.7$~$M_{\rm Jup}$, Gaia-4b is the first confirmed exoplanet astrometrically detected by {\it Gaia}.\footnote{Gaia-3b (a.k.a.\ HIP~66074b, a super-Jupiter on a 300d period orbit) was originally announced as the first astrometric {\it Gaia} candidate by \citealt{Sozzetti2023}. However, a subsequent internal review of the {\it Gaia} pipeline showed that a software bug affected the target’s time series, leading to a false-positive orbital solution in {\it Gaia} DR3 \citep{Pinamonti2025}.
}

At the time of writing, the rest of the orbital solutions in the planetary-mass range are associated to exoplanet candidates\footnote{The complete list of {\it Gaia} exoplanet candidates is available at 
\href{https://www.cosmos.esa.int/web/gaia/exoplanets}{{\tt https://www.cosmos.esa.int/web/gaia/exoplanets}}}, 
although efforts are being made to confirm or refute them \citep[e.g., see][]{Stefansson2025}.

\subsection{Periodogram}

The Lomb-Scargle (LS) periodogram is a commonly used tool to look for periodicities in radial velocity time series, based on Fourier decomposition principles. \cite{Anglada-Escude2010} developed a modified version of the LS periodogram to search for periodic signals in two-dimensional astrometric data. Consider the simplest case of a periodic signal due to a single planet in a circular orbit ($e=0$). A least-squares fitting using the basic astrometric model (the null hypothesis) with $k_0=5$ parameters ($\alpha_0$, $\delta_0$, $\mu_\alpha$, $\mu_\delta$, and $\pi$) yields a reduced chi-square of $\chi_0^2$. Including a Keplerian orbit of period $P$ in the model will improve the $\chi^2$ of the fit. The least-squares fitting of the astrometric plus Keplerian model with $k_P=8$ parameters ($\alpha_0$, $\delta_0$, $\mu_\alpha$, $\mu_\delta$, $\pi$, $a_1$, $\Omega$, and $i$; $P$ is fixed) yields $\chi_P^2$. 
The periodiogram power, $z$, is obtained by comparing the reduced chi-squared of the model that includes a planet to the reduced chi-squared of the null hypothesis,

\begin{equation}\label{eq:periodogram}
 z (P) = \frac{(\chi_0^2 - \chi_P^2)/(k_P - k_0 )}{\chi_P^2 / ( N_{\rm data} - k_P ) } 
\end{equation}

\noindent where $N_{\rm data}$ is the number of data points. For astrometric two-dimensional data, $N_{\rm data} = 2\times N_{\rm obs}$, where $N_{\rm obs}$ is the number of observed epochs. Thus, $z(P)$ measures how much the $\chi^2$ of the fit improves by adding a Keplerian orbit in the model.

To search for a planetary signal, $z(P)$ is evaluated for a wide range of periods. The significance of a power peak in the periodogram is determined by the False Alarm Probability (FAP), which tells the probability that the observed power would appear purely due to noise  \citep{Cumming2004}. 

The periodogram is a useful tool to localize significant signals and narrow down the period search window in the astrometric fits. It can be generalized to search for periodic signals due to several companions. For instance, if a signal has been already detected, the periodogram compares the $\chi^2$ of the fit of a one-companion model to the fit of a two-companion model \citep{Curiel2020}. In the more general case, the periodogram power is given by,

\begin{equation}
 z (P) = \frac{(\chi_{N_P}^2 - \chi_{N_{P+1}}^2)/(k_{N_{P+1}} - k_{N_P} )}{\chi_{N_{P+1}}^2 / ( N_{\rm data} - k_{N_{P+1}} ) },  
\end{equation}

\noindent where $\chi_P^2$ and $\chi_{P+1}^2$ are the $\chi^2$ for the model with $N_P$ and $N_{P+1}$ planets, respectively, and $k_{N_{P}}$ and $k_{N_{P+1}}$ the number of free parameters in each model.

\subsection{Proper Motion Anomaly}

Other astrometric exoplanet discoveries have searched for anomalies in the stellar proper motions as an indicator of the presence of hidden orbiting companions. This technique consists of measuring the difference between the \textsc{Hipparcos}--{\it Gaia} mean proper motion and the {\it Gaia} proper motion, which is referred as the proper motion anomaly or acceleration \citep[][]{Brandt2018,Kervella2019}. Thanks to the long time baseline of $\approx$25 years between Hipparcos and {\it Gaia} observations, it is possible to derive the long-term proper motion of nearby stars with a high accuracy. The proper motion anomaly, $\Delta\mu = \mu_{{Gaia}} - \mu_{{\rm Hipparcos}-Gaia}$, can be used to constrain the mass of possible unseen companions. Assuming  a circular and  “face-on” orbit, $\Delta\mu$ is linked to the orbital velocity $v$ via,

\begin{equation}\label{eq:pma}
    v = {d} {\Delta\mu} \approx \sqrt{ \frac{G m_2^2}{m_1  r} },
\end{equation}

\noindent where where $m_1$ is the mass of the primary star, $m_2$ the mass of the companion, $d$ the distance to the star and $G$ the  Gravitational constant. This equation can be rewritten as, 

\begin{equation}
    \frac{m_2}{\sqrt r} \approx \sqrt{ \frac{m_1}{G} } \left( \frac{\Delta \mu [\rm{ mas~yr^{-1}}]}{\varpi [\rm{ mas}]} \times 4740.47 \right) .
\end{equation}

Therefore, if $m_1$ is known, it is possible to estimate the mass $m_2$ of a companion as a function of the
separation from the host star  \citep{Kervella2019}. 
Using this technique, \cite{Kervella2022} identified more than 12000 bound candidate companions in the \textsc{Hipparcos}  catalog\footnote{\url{https://www.cosmos.esa.int/web/hipparcos/catalogues}}, a fraction of which are in the substellar and even in the planetary-mass regime.   
For some of these planet candidates, combined RV and astrometric analyses has been used to break the degeneracy between companion mass and orbital parameters \citep[e.g.][]{Feng2022}. 

The reader interested in an inventory of stars with significant accelerations can consult the \textsc{Hipparcos}–{\it Gaia} Catalog of accelerations \citep{Brandt2018,Brandt2021}. Long-term proper motions are also available in the Tycho--{\it Gaia} astrometric solution (TGAS), which combines the \textsc{Hipparcos} Tycho--2 catalogue  \citep{Hog2000} and {\it Gaia} DR1 astrometry.

\subsection{Planets in Binary Systems } 

Astrometry can unveil the presence of exoplanets in binary systems from both relative and absolute astrometric data. 

\subsubsection{S-type orbits}

Let us first consider  a satellite type (S-type) architecture \citep{Dvorak1982}, where the exoplanet orbits only one of the stars of the binary system. 
In the case of relative astrometry, the positions of the stellar companion are measured with respect to the primary star. The presence of an additional unseen inner companion is inferred from deviations from a single Keplerian orbital model. 
If the binary orbit is significantly larger than the inner orbit, the relative positions can be fitted with a double Keplerian orbital model, which consists of the sum of the binary orbit plus the orbit of the inner companion \citep{Gardner2022}. The inner orbital elements will describe the motion of the primary star about the barycenter of the inner orbit, this is, the star wobble induced by the planetary companion. The outer orbital elements will describe the binary orbit.

In the case of absolute astrometry, the  positions of the primary star are given by,

\begin{equation}\label{eq:planet_binary_1_ra}
\alpha^A(t) = \alpha_0 + \mu_\alpha \cos\delta \cdot (t - t_0) + \varpi \cdot f_\alpha(t) + G_\alpha^{A}(t) + G_\alpha^{A,p}(t), 
\end{equation}
\begin{equation}\label{eq:planet_binary_1_dec}
\delta^A(t) = \delta_0 + \mu_\delta \cdot (t - t_0) + \varpi \cdot f_\delta(t) + G_\delta^{A}(t) + G_\delta^{A,p}(t),
\end{equation}

\noindent while the positions of the secondary can be written as,

\begin{equation}\label{eq:planet_binary_2_ra}
\alpha^B(t) = \alpha_0 + \mu_\alpha \cos\delta \cdot (t - t_0) + \varpi \cdot f_\alpha(t) - \left( \frac{a_2}{a_1}\right)\cdot G_\alpha^{A}(t), 
\end{equation}
\begin{equation}\label{eq:planet_binary_2_dec}
\delta^B(t) = \delta_0 + \mu_\delta \cdot (t - t_0) + \varpi \cdot f_\delta(t) - \left(\frac{a_2}{a_1}\right)\cdot G_\delta^{A}(t),
\end{equation}

\noindent where $a_1$ and $a_2$ are the semi-major axis of the primary and secondary, respectively, and $(\alpha_0,\delta_0)$, $(\mu_\alpha,\mu_\delta)$ and $\varpi$ are the reference position, proper motion and parallax of the system's barycenter, respectively. The terms ($G_\alpha^{A}(t)$,~$G_\delta^{A}(t)$) and  ($G_\alpha^{A,p}(t)$,~$G_\delta^{A,p}(t)$) are the components of the Keplerian orbit of the primary star around the barycenter, induced by the outer and inner companion, respectively. 
Even in the case that only the primary star is detected, it is possible to simultaneously fit the Keplerian orbit of the binary, the orbit of the inner companion, as well as the parallax and proper motion of the barycenter (eq.\ \ref{eq:planet_binary_1_ra}--\ref{eq:planet_binary_1_dec}). 
In this case the model will have 5 astrometric parameters plus $7 \times 2$ orbital parameters. If both the primary and the secondary  are detected, then equations (\ref{eq:planet_binary_1_ra})--(\ref{eq:planet_binary_2_dec}) are fitted simultaneously to the astrometric data of the two stars. In this case there is an additional parameter, namely $a_2$ or $q=a_1/a_2$, that appears in the model for the secondary star, resulting in 20 total parameters.

If additional relative positions of the binary orbit are available, for instance from optical/infrared relative astrometry, these can be combined with the absolute astrometry to simultaneously fit the relative orbit of the binary and the absolute astrometric data of the primary and the secondary stars, as well as the orbit of the inner companion \citep{Curiel2022}. 

\begin{figure}[!th]
\begin{center}
 \includegraphics[width=0.7\textwidth,angle=0]{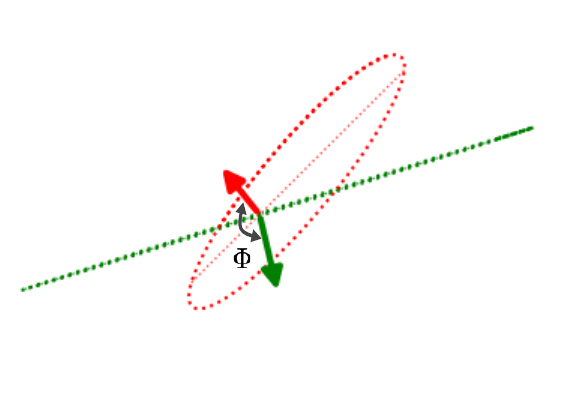}
\caption{3D depiction of the mutual inclination angle ($\Phi$) between two orbital planes. The green orbit is seen edge-on. The arrows mark the direction of the angular momentum vectors, which are perpendicular to the orbital planes. The mutual inclination angle is the angle between these two vectors, measured in the plane that contains them. }
\label{fig:inclination-angle}
\end{center}
\end{figure}

In the case that the number and cadence of the observations cover properly both the binary and the planet orbits, all the Keplerian elements can be obtained from the astrometric fits. In this case, the mutual inclination angle between the two orbital planes, $\Phi$ (see Fig. \ref{fig:inclination-angle}), can be estimated from the measurements of the inclination angles and the position angles of the ascending node, through the equation \citep{Muterspaugh2006},

\begin{equation}
 \cos\Phi  = \cos{i_{Ab}} \cos{i_{AB}}  + \sin{i_{Ab}} \sin{i_{AB}}( \cos{\Omega_{Ab} - \Omega_{AB}}). 
\end{equation}

\noindent where $i_{Ab}$ are the inclination angles of the planet and binary orbit, respectively\footnote{The subindex ``b''  is used to designate a planetary-mass companion around the primary and  ``B'' to the stellar companion.}, and $\Omega_{Ab}$ and $\Omega_{AB}$ are the position angles of the ascending node of these orbits (see Fig. \ref{fig:sky-orbit} and \ref{fig:3D-orbit} for the definition of these angles). 

While the mutual inclinations between planetary and stellar orbits can be addressed statistically \citep[e.g.,][]{Dupuy2022}, the true three-dimensional orbital architecture of a binary-planetary system can only be recovered if the orbital angles are well constrained. Indeed, the three angles of a Keplerian orbit (argument of periastron, position angle of the ascending node, and inclination angle) can be obtained directly with the astrometry technique, subject to the ambiguity of 180 degrees in the position angle of the ascending node, but this can be solved with the addition of radial velocity information.

\subsubsection{P-type orbits}

\begin{figure}[!th]
\begin{center}
 \includegraphics[width=0.6\textwidth,angle=0]{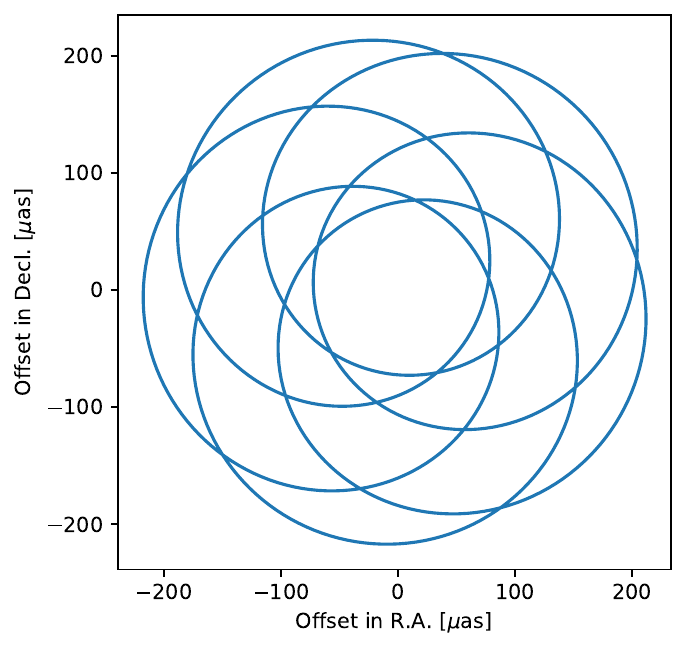}
\caption{Simulated photocenter motion of a circumbinary planet with mass of $7.9 M_{\rm Jup}$ and period of 240 days. The binary masses are 0.96 and  $0.94 M_\odot$. The binary period is 40 days.  Adapted from \cite{Sahlmann2015}.}
\label{fig:rosette}
\end{center}
\end{figure}

In a planetary-type (P-type) o circumbinary orbit the exoplanet orbits both stars of the binary system. In this case, it is possible to model the planet motion as an orbital motion around the center of mass of the binary.  
Here, one should consider that the binary may be unresolved, such that only the binary photocenter is observed. 

The photocenter is defined as the flux-weighted mean position of the two stars \citep{Marcussen2023},

\begin{equation}
a_0 = \left| \frac{F_1 a_1 - F_2 a_2}{ F_1 + F_2 } \right|
\end{equation}

where $F_1$ and $F_2$ are the fluxes of the primary and secondary star, respectively, and $a_1$ and $a_2$ their semi-major axes.  When the flux of the secondary is too faint ($F_2\approx0$), the photocenter is on the primary star ($a_0 \approx a_1$). In other words, the motion of the photocenter closely follows the brightest star. On the contrary,  in the case where the two stars have equal fluxes ($F_2=F_1$) and equal mass ($M_2=M_1$), then the photocenter has no visible orbital motion (it remains in the barycenter of the system). The photocenter motion, with amplitude $a_0$, is sometimes referred as the astrometric binary signal \citep[e.g.,][]{Sahlmann2015}.

A  circumbinary planet  induces an additional astrometric signature, resulting in a combined motion that has a rosette shape (Figure \ref{fig:rosette}).
The total motion of the binary system will have three contributions \citep{Sahlmann2015}: (a) parallax and proper motion of the binary, (b) the binary photocenter motion, with amplitude $a_{0}$ (i.e., the astrometric binary signal), and (c) the orbital motion of the photocenter due to the orbiting planet, with amplitude  $a_{\rm P}$. Accounting for these contributions, the model will have 19 free parameters: 5 astrometric (two reference positions, two proper motions and parallax), plus 7$\times$2 Keplerian parameters (for the binary and planetary orbits).

\subsection{Infrared astrometry }

So far, only one exoplanet has been discovered solely through infrared astrometry \citep{Muterspaugh2010V}. This planet is orbiting around one of the components of HD~176051, a nearby ($d\sim15$~pc) binary system composed of two low-mass stars (1.07 and 0.71~$M_\odot$, for the primary and secondary stars, respectively) and orbital period of 61.41~yr. Assuming the planet is orbiting the secondary star,  \citep{Muterspaugh2010V} derived a planet mass of $1.5\pm0.3~M_{\rm Jup}$. If the planet is orbiting the primary star, the planet mass is 2.26~$M_{\rm Jup}$, although the secondary star is more favoured as the exoplanet host.

Infrared interferometry has the potential to detect wobbles in the binary motion induced by short-period inner substellar companions. This was demonstrated by \cite{Muterspaugh2010} through the implementation of the Palomar High-precision Astrometric Search for Exoplanet Systems (PHASES) program, using the Palomar Testbed Interferometer (PTI), which was located on Palomar Mountain near San Diego, California. PHASES measured the differential astrometry (the angle between two stars) of 51 subarcsecond binary systems \citep{Muterspaugh2010V}. By simultaneously observing a pair of stars (an observing mode called dual-field), the effect of atmospheric turbulence cancels out, which makes very high precision ``narrow-angle'' astrometry possible. PHASES achieved an astrometric accuracy of $\sim 100~\mu$as, and found evidence of substellar companions in six binaries, including HD~176051.

The early-generation projects ASTrometric and phase-Referencing Astronomy (ASTRA; \citealt{Woillez2010}) at the Keck Observatory and the Phase-Referenced Imaging and Microarcsecond Astrometry (PRIMA; \citealt{Delplancke2006}) at the Very Large Telescope astronomical facility, were designated to use  world-class telescopes for infrared long-baseline interferometry.  By aiming to achieve astrometric precision of  $\sim10~\mu$as, one of their primary goals was the detection and characterization of exoplanets. However, technical challenges that prevented reaching the required level of precision, along with funding constraints, ultimately led to the discontinuation of these instruments.

The Center for High Angular Resolution Astronomy (CHARA) Array is an optical/near-IR interferometer with baselines up to 330~m \citep{tenBrummelaar2005}. CHARA measures the differential astrometry between two components of a binary system within the interferometric field-of-view. The Michigan InfraRed Combiner-Exeter (MIRC-X) instrument is used to combine all six telescopes available at CHARA \citep{Anugu2020}. 

GRAVITY \citep{GRAVITYCollaboration2017}, installed at the Very Large Telescope Interferometer (VLTI), is an interferometric instrument operating in the near infrared. 
It combines  four telescope beams, reaching baselines of up to 200~m, and can operate in dual-field mode to obtain high-precision relative astrometry.

Both CHARA/MIRC-X and VLTI/GRAVITY routinely achieve an accuracy of a few 10~$\mu$as \citep{Eisenhauer2023}. The two instruments were used by \cite{Gardner2022} to conduct the ARrangement for Micro-Arcsecond Diﬀerential Astrometry (ARMADA) survey with the aim of detecting previously unseen companions in intermediate mass binary systems down to the planetary mass regime. The survey discovered nine companions that had not been detected before by any method, although none have planetary masses.

\cite{GRAVITYCollaboration2024} reported a Neptune-mass candidate planet orbiting one of the two M dwarfs in the nearest visual binary system, GJ~65AB. The infered planet mass is $36\pm7$ or $39\pm7~M_{\oplus}$ depending on whether it is associated with the secondary or the primary star, respectively. The astrometric signature induced by the planet is $\sim 100$~mas, while the GRAVITY measurements had a mean accuracy of 50--60~$\mu$as. This indicates a  $2\sigma$ detection of the astrometric signal, and that additional observations are required to confirm its planetary nature.

Although this technique has yielded only a small number of confirmed or candidate exoplanets detected through the astrometric signature of the host star, the ultra-high astrometric accuracy of infrared interferometers has been crucial for determining precise orbits of directly imaged exoplanets -- such as $\beta$~Pic~c \citep{Nowak2020} and~HD 206893c \citep{Hinkley2023} -- thereby enabling accurate measurements of their masses. We refer the reader to {Chapter 5} for a comprehensive discussion of the capabilities of infrared interferometry  for directly imaging exoplanets.

\section{Demographics of the detected planets}

\begin{table}
\scriptsize
\caption{Confirmed exoplanets and brown dwarfs discovered with the astrometry technique.}
\label{tab:exoplanets} 
\centering 
{
\begin{tabular}{l c c c c c c c c}  
\hline\hline  
Name  &  $m_p$  & $a_p$  &  $P$ & $M_\star$ &  Spectral &Discovery \\
& (M$_{\rm Jup}$)& (au) & (days) & (M$_\odot$) &  Type & \\
\hline 
TVLM 513-46546 b & 0.38 & 0.3 & 220.0 & 0.08 & M8.5V & 2020 \\
HD 92945 c & 0.7 & 14.64 & -- & 0.86 & K1V & 2024 \\
HD 176051 Bb & 1.5 & 1.76 & 1016.0 & 0.9 & K1V & 2010 \\
HIP 1481 Ab & 1.6 & 14.17 & -- & 1.17 & F8V & 2022 \\
HIP 560 b & 1.7 & 16.57 & -- & 1.45 & F2V & 2024 \\
GJ 896 Ab & 2.26 & 0.64282 & 284.39 & 0.44 & M4Ve & 2022 \\
HIP 96334 b & 3.7 & 6.15 & -- & 1.02 & G3V & 2022 \\
HD 155555 (AB)b & 3.8 & 7.28 & -- & 0.86 & G5+K1 & 2024 \\
HD 14082 Bb & 5.4 & 5.74 & -- & 1.09 & G1V & 2024 \\
AB Pic c & 5.75 & 3.571 & -- & -- & K2 V & 2022 \\
HIP 47110 b & 6.0 & 16.5 & -- & 0.96 & G5V & 2024 \\
WISE J1355-8258 b & 9.0 & -- & -- & 0.01 & sdL5 & 2018 \\
2MASS J1155-7919 b & 10.0 & 582.0 & -- & -- & M3 & 2020 \\
HIP 11696 b & 10.0 & 15.25 & -- & 1.32 & F5V & 2024 \\
2MASS J0249-0557 (AB)b & 11.6 & -- & -- & -- & -- & 2018 \\
Gaia-4 b & 11.8 & -- & 571.3 & 0.64 & -- & 2024 \\
GJ 2030 Ac & 12.934 & 16.752 & 25635.0 & 0.96 & G5IV & 2022 \\
2MASS J0434+1722 b & 13.0 & 1.7 & -- & 0.18 & M4.25 & 2025 \\
\hline 
HIP 54515 b & 17.7 & 24.8 & 32507.0 & 1.88 & A5V & 2025 \\
Gaia-6 b & 19.8 & 4.85 & 3420.05 & 1.21 & F8V & 2024 \\
Gaia-5 b & 20.93 & -- & 358.57 & 0.36 & -- & 2024 \\
MHO 6 c & 26.0 & 1.83 & -- & 0.19 & M5 & 2025 \\
2MASS J1610-3922 b & 27.0 & 1.6 & -- & 0.2 & M4.5 & 2025 \\
LSPM J1831+4213 b & 32.09 & 0.468 & 324.05 & 0.11 & M7V & 2023 \\
GJ 229 Bb & 33.4 & 0.0424 & 12.1346 & 0.54 & M1V & 2021 \\
PDS 70 e & 34.0 & 4.5 & -- & 0.76 & K7 & 2025 \\
72 Peg Bb & 35.0 & 3.46 & 1539.0 & 2.0 & K5III & 2010 \\
DoAr 21 (AB)b & 35.6 & 1.55 & 450.88 & 2.04 & K0 + M3 & 2019 \\
Sz 76 b & 36.0 & 1.6 & -- & 0.22 & M3.2 & 2025 \\
HIP 58289 b & 40.0 & 23.6 & 45625.0 & 0.78 & K2V & 2025 \\
Sz 84 b & 40.0 & 1.25 & -- & 0.17 & M5.0e & 2025 \\
2MASS J0249-0557B & 44.0 & -- & -- & -- & -- & 2018 \\
DoAr 21 (AB)c & 44.0 & 2.8 & 1013.5 & 2.04 & K0 + M3 & 2019 \\
HIP 117179 b & 44.197 & 0.788 & 247.98 & 1.02 & G3V & 2023 \\
LP 769-9 b & 44.8 & 0.852 & 339.6 & 0.67 & K7V & 2024 \\
LSPM J1657+2448 b & 46.195 & 1.198 & 1182.34 & 0.12 & M7V & 2023 \\
TYC 8321-266-1 b & 46.226 & 1.157 & 413.89 & 1.16 & F8V & 2023 \\
NLTT 35024 b & 46.424 & 0.608 & 412.73 & 0.11 & M5.5V & 2023 \\
LP 769-41 b & 47.2 & 0.448 & 150.8 & 0.48 & M2V & 2024 \\
2MASS J0249-0557A & 48.0 & -- & -- & -- & -- & 2018 \\
HD 104289 b & 49.483 & 2.42 & 1233.33 & 1.21 & F8IV-V & 2023 \\
2MASS J19262435-1045398 b & 49.8 & 0.431 & 175.8 & 0.3 & M4V & 2024 \\
L 499-75 b & 50.0 & 0.612 & 240.3 & 0.48 & M2.5V & 2024 \\
Gaia DR3 4224582166524567040 b & 50.8 & 0.912 & 413.7 & 0.53 & M1V & 2024 \\
SIPS J2049-2800 b & 52.0 & 0.822 & 641.1 & 0.13 & M6.5V & 2024 \\
DE0630-18 (bc) & 53.0 & -- & -- & -- & -- & 2020 \\
2MASS J12114960+7752445 b & 54.3 & 1.15 & 651.2 & 0.43 & M2V & 2024 \\
G 165-52 b & 54.364 & 0.479 & 156.22 & 0.55 & M0V & 2023 \\
HD 340935 b & 58.0 & 2.22 & 1176.2 & 1.01 & G2V & 2024 \\
DE0630-18 a & 58.0 & -- & -- & -- & -- & 2008 \\
SDSS J080531+481233 A & 60.0 & -- & -- & -- & -- & 2016 \\
HIP 71618 b & 60.0 & 11.1 & 9350.0 & 2.05 & A1V & 2025 \\
Sz 100 b & 61.0 & 1.1 & -- & 0.14 & M5 & 2025 \\
UCAC2 9182345 b & 61.0 & 0.455 & 130.3 & 0.69 & K6V & 2024 \\
TYC 9255-929-1 b & 61.629 & 0.836 & 298.47 & 0.88 & K0V & 2023 \\
GJ 802 (AB)b & 63.0 & 1.32 & 1104.0 & 0.18 & M5Ve+M5Ve & 2005 \\
TYC 3873-761-1 b & 63.302 & 1.255 & 526.19 & 0.92 & G7V & 2023 \\
HD 115517 b & 64.475 & 1.3347 & 439.49 & 1.0 & G3IV-V & 2023 \\
TYC 3056-264-1 b & 64.727 & 1.25 & 564.96 & 0.77 & K2V & 2023 \\
UCAC4 302-050985 b & 65.153 & 0.647 & 276.51 & 0.63 & M0V & 2023 \\
HD 156312 Bb & 66.519 & 0.744 & 238.43 & 0.99 & G5V & 2023 \\
TYC 7922-716-1 b & 67.0 & 0.818 & 288.7 & 0.86 & K0V & 2024 \\
CD-41 1115 b & 68.0 & 1.095 & 406.5 & 0.99 & G5V & 2024 \\
GSC 04516-00523 b & 68.205 & 0.352 & 87.7 & 0.68 & K7V & 2023 \\
HIP 60321 b & 68.26 & 1.15 & 530.17 & 0.64 & M0V & 2023 \\
HIP 75202 Ab & 69.0 & 1.316 & 591.46 & 0.82 & K1V & 2023 \\
\hline 
\end{tabular}
}
\end{table}

\begin{figure}[!th]
\begin{center}
 \includegraphics[width=0.7\textwidth,angle=0]{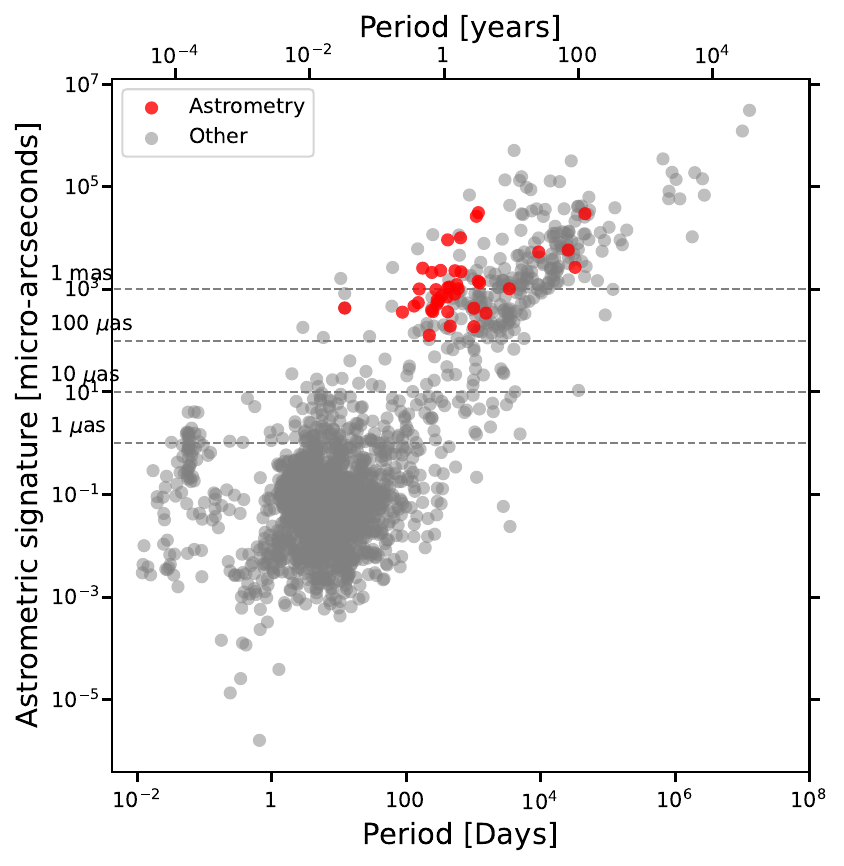}
\caption{Astrometric signature of confirmed exoplanets as a function of orbital period. The red dots correspond to planets discovered with the astrometry technique, and gray dots are planets discovered with any other method.  }
\label{fig:as-vs-p}
\end{center}
\end{figure}

\begin{figure}[!th]
	\begin{center}
		\includegraphics[width=0.6\textwidth,angle=0]{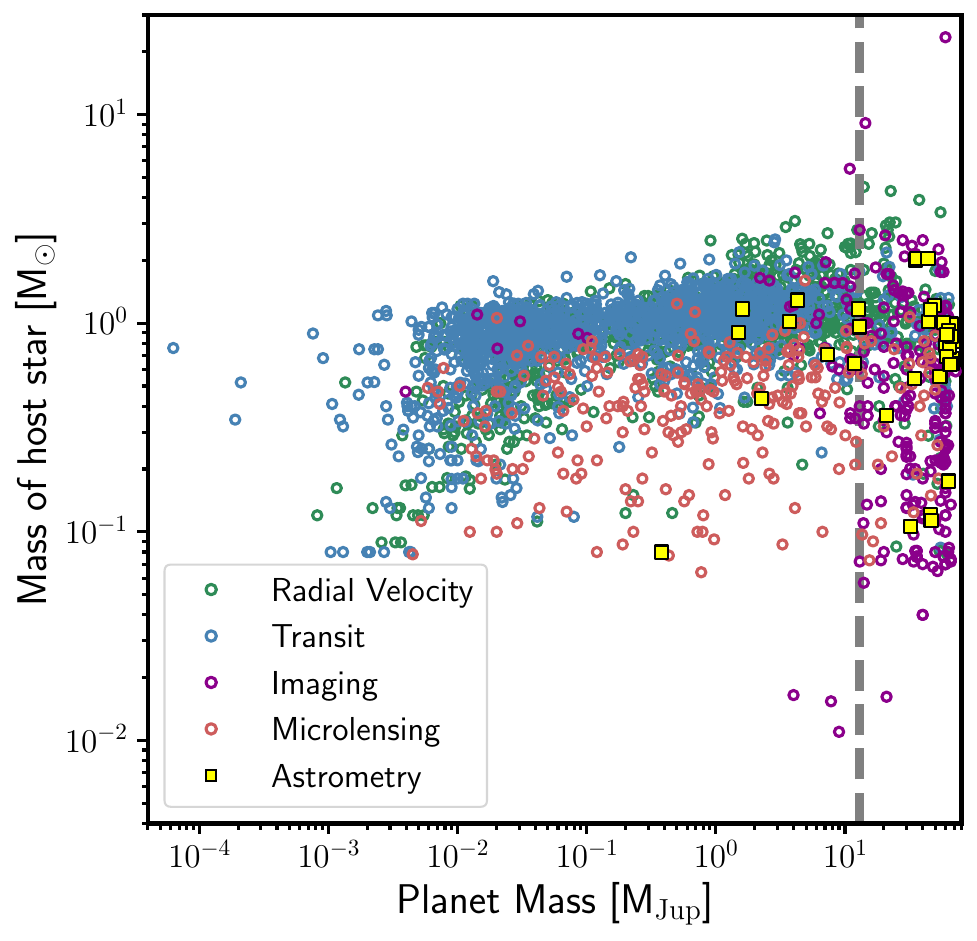}
		\includegraphics[width=0.6\textwidth,angle=0]{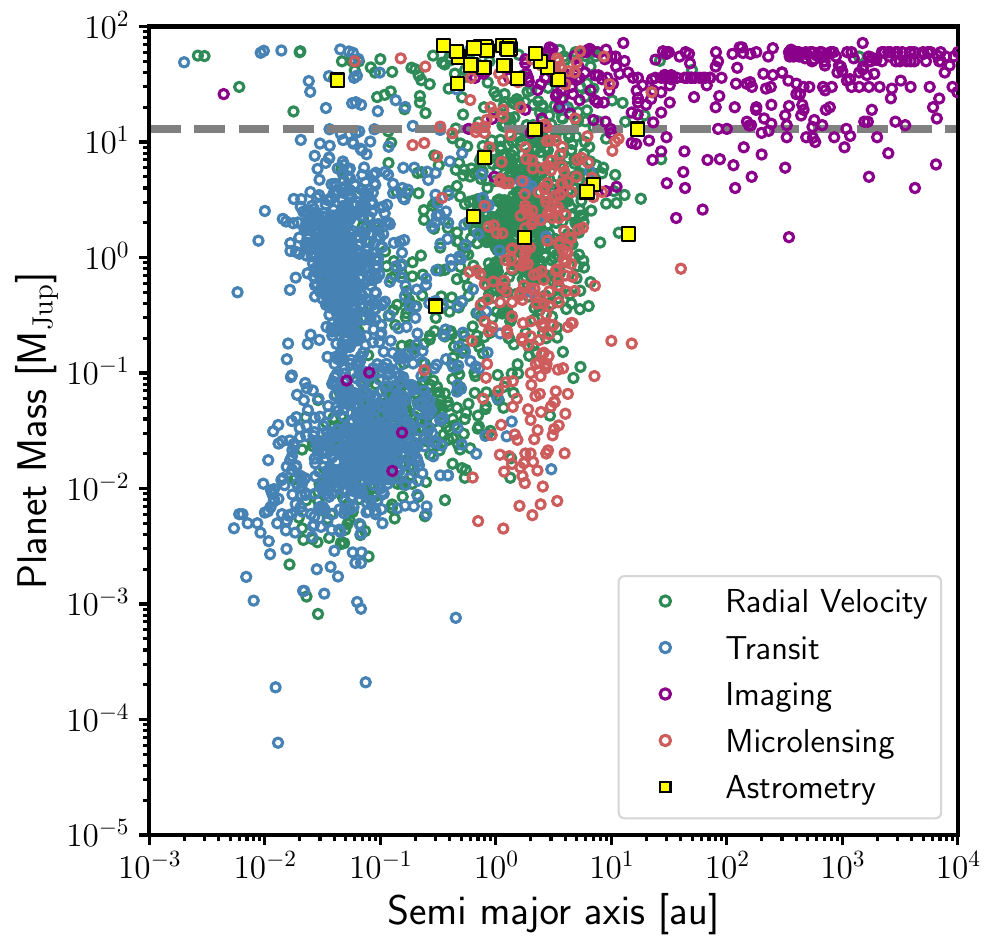}
		\caption{Distributions of planetary mass, mass of the host star and semi-major axis of all confirmed exoplanets listed at \href{https://exoplanet.eu/home/}{{\tt exoplanet.eu}}. The planets are color-coded according to method used for their discovery. The dashed gray line marks the deuterium-burning mass limit of 13 $M_{\rm Jup}$. }
		\label{fig:demographics}
	\end{center}
\end{figure}

All confirmed exoplanets discovered with the astrometry technique are listed in Table \ref{tab:exoplanets}. The exoplanets were retrieved from the Extrasolar Planets Encyclopedia (\href{https://exoplanet.eu/home/}{{\tt exoplanet.eu}}) and are ordered by mass. At the time of writing, only 18 of these objects have masses below the deuterium-burning mass limit of 13~$M_{\rm Jup}$. 
However, it should be noted that the mass limit used for a planetary-mass object to be included in this catalog is 60~$M_{\rm Jup}$ plus one sigma. The choice of 60~$M_{\rm Jup}$ is based on studies of the mass–density relationship for low-mass planets and stars that shows a change in the slope  
at 60~$M_{\rm Jup}$ \citep{Hatzes2015}.
Twelve of the planets  with $M_p<13~M_{\rm Jup}$ listed in Table \ref{tab:exoplanets} were discovered from measurements of the proper motion anomaly  (e.g., HIP 1481 Ab, HIP 96334 b and GJ 2030 Ac), one was found from relative astrometry observations of a binary system (HD 176051 Bb),  and three were discovered using absolute astrometric data from  VLBI observations or {\it Gaia} (TVLM 513-46546~b, GJ 896~Ab and Gaia-4~b).

Figure \ref{fig:as-vs-p} shows the astrometric signature as a function of the orbital period for all planetary-mass objects reported in The Extrasolar Planets Encyclopedia. To compute the astrometric signature I used Eq.\ (\ref{eq:asignature}) if $M_p$ and $a_p$ are both listed in the catalog. If $a_p$ is not listed, I use 

\begin{equation}
a_{\rm p} = M_\star \left( \frac{P} {M_\star + M_p} \right)^{2 /3}  
\end{equation}

\noindent where $a_{\rm p}$ is in au, $P$ in years and the masses are in Solar masses. The astrometric signature was not computed if the mass of the host star is not listed in the catalogue.

The astrometrically detected exoplanets have astrometric signatures above 100~$\mu$as, as  expected given the astrometric precision of current telescopes. Below this limit, the detection of signals at high confidence  are difficult since they would require an astrometric accuracy better than $\approx30$~$\mu$as. In addition, a good coverage of the planet orbit is necessary to constrain well the Keplerian elements, in particular the eccentricity and the argument of periastron passage, thus requiring significant investments in terms of observing resources.

As astrometric missions continue to observe, the ability to detect planetary signatures and reconstruct exoplanetary orbits will improve. Estimations on the number of expected detections suggest that, for its nominal 5--yr mission, {\it Gaia} will discover some $\sim 21,000 \pm 6,000$ massive (1--15~$M_{\rm Jup}$) and long-period planets out to distances of $\sim$500~pc \citep{Perryman2014}. Of these, about 1,000 are expected to be orbiting M-dwarfs within 100~pc. The reason of such a difference is  related to a greater volume of space that is probed by a larger distance, which increases the number of stars for which planets can be astrometrically detected. Mass constraints from astrometric accelerations  will also improve, as the precision of the acceleration measurement scales as $t^{5/2}$, where $t$ is the observing time baseline \citep{Brandt2024}.

The advent of forthcoming radio interferometric arrays will have a big impact on the field of exoplanet characterization.
For instance, the next generation Very Large Array (ngVLA) is a new facility planned to achieve (sub)milliarcsecond angular resolutions at centimeter wavelengths. The ngVLA will deliver astrometric accuracies of 1 $\mu$as \citep{Reid2018}, which will be possible thanks to the enhanced sensitivity of the array and improvements to the calibration techniques \citep{Rioja2020}. This ultra-high astrometric accuracy represents an increase by a factor of 10 with respect to the astrometry delivered by current instrumentation. An analysis similar to that presented in Fig.\ \ref{fig:signature} suggests that, with this level of accuracy, the ngVLA should be able to detect the signatures of Neptune-like planets and Super-Earths around nearby stars ($d\sim10$ pc) and Jupiter-like planets around stars located in nearby star-forming regions ($d\sim140$ pc), although it is not clear how many of these planets can be detected. 

\subsection{Host-star demographics}

According to Equation (\ref{eq:asignature}), the astrometric signature scales with the ratio of the planet mass to the host star mass. It also scales with the semi-major axis (or period) of the planet's orbit and the distance to the star. Therefore, massive planets ($\gtrsim 0.3 M_{\rm Jup}$) in long periods ($\gtrsim$ 1 year) associated to nearby ($d\lesssim$10 pc) low-mass stars ($M \lesssim 0.1 M_\odot$) are the easiest to detect with this technique. As mentioned above, if the astrometric accuracy of current instruments improves by a factor of 10, it would be possible to detect Neptune-like planets and Super-Earths around this kind of stars. 

Because astrometric measurements require high precision, which in turn requires the star’s emission to be compact at the resolution of the observations, the astrometry method cannot be applied to all types of stars. For instance, the large angular sizes of giant stars  would result in less accurate astrometric measurements, which will hamper the detection of the tiny reflex motion caused by orbiting planets. The high mass of OB stars also reduces the magnitude of the reflex motion, which makes harder to detect the astrometric signature in these stars.  Table \ref{tab:exoplanets} gives the spectral types of the stars hosting all astrometrically detected planetary-mass objects, where it can be seen that the hosts have late-F to late-M spectral types, and the majority are main-sequence stars.     

The observing wavelength also imposes some limitations on the applicability of the astrometry method. For instance, at optical wavelengths, interstellar and circumstellar dust both contribute to making young stars nearly invisible. Observations at infrared and radio wavelengths offer a good alternative for astrometric planet searches in such stars.  

Very low mass stars and brown dwarfs are also intrinsically faint in the optical. These stars show broad molecular spectral features and strong stellar activity. This is a challenge for exoplanet searches with the transit and radial velocity techniques.  The RV scatter due to stellar activity can be larger than the RV signal induced  by planets \citep{Saar1997}. As a consequence, RV surveys generally avoid active M dwarfs {{(see Chapter 2)}}. Activity-related variability also affects transit light curves {{(see Chapter 3)}}. For instance, periodic variations caused by starspots can generate confounding signals \citep{Berta2012}. Low-mass young stars (a few tens of Myr) show even stronger levels of stellar activity due to enhanced magnetic fields. This makes the detection of planets around young stars very challenging for the RV and transit technique. Magnetically active stars are, on the other hand, strong radio emitters. Thus, radio astrometry can be used to search for exoplanets not accessible by other techniques.

Currently, only a few Jupiter-like planets have been found orbiting stars with masses $M\lesssim 0.1 M_\odot$ (see top panel of Fig.\ \ref{fig:demographics}), while  several brown dwarfs with masses $\approx 20-60~M_{\rm Jup}$ have been revealed around the same kind of stars by direct imaging. Given the challenges mentioned above, radio astrometry holds particular promise in this regard.

The distribution of planet mass versus semi-major axis of all confirmed exoplanets is shown in the bottom panel of Figure \ref{fig:demographics}. The detection method is indicated by the color of the plotting symbols. Planets discovered with astrometry seem to occupy the same locus in this plot as the locus occupied by planets detected by radial velocity. However, as was discussed before, astrometry is well suited for unveiling giant planets associated with very low-mass stars. Thus, it can fill in the parameter space where few detections have been obtained so far. Furthermore, planned astrometric facilities, like next-generation radio interferometers, will provide the necessary sensitivity and astrometric accuracy to detect low-mass planets, which will complement the exoplanet demographics probed by other observational techniques. 

Since astrometry is most sensitive to massive planets with large orbits, a substantial sample of astrometrically detected exoplanets can test theoretical models of giant planet formation. 
Planetary formation occurs in the circumstellar disks surrounding young stars. However, the classical ``Core Accretion'' scenario cannot explain the formation of gas giants orbiting low-mass stars because this mechanism requires a massive ($\sim 10~M_\oplus$) core to form before the gaseous disk disperses \citep{Mizuno1980,Laughlin2004}.  The ``Gravitational Instability'' scenario is another mechanism proposed to explain the formation of gas giants by the fragmentation and collapse of the circumstellar disk \citep{Boss1997}.  Numerical simulations have found that giant planets can form on wide orbits  on rapid timescales ($\sim 20-30$ kyrs) even around brown dwarfs \citep{Mercer2020}.
However, once formed, planets  evolve dynamically by migrating inward or outward due to disk perturbations and the presence of other planets or stellar companions, thereby altering the original architecture of the systems. Therefore, the architecture of young planetary systems, which can be studied by astrometry, will provide key information on planetary formation. On the other hand, 
uncovering the architecture of evolved systems will help to understand the dynamical evolution of planets.  For instance, {\it Gaia} is expected to detect planets with masses above $\sim3~M_{\rm Jup}$ and at a few au around the nearest and brightest white dwarfs (WD; \citealt{Silvotti2011,Sanderson2022}). These discoveries will be crucial to investigate processes like dynamical scattering and survival of common envelope evolution happening in the post-main sequence, and probe the formation of second or third generation planets \citep{Sanderson2022}. The potential of {\it Gaia} to find exoplanets around  white dwarfs has been demonstrated by the discovery of WD~0141--675~b, a companion candidate in the planetary-mass range. This companion has a period of 33.65$\pm$0.05 days and with a WD mass of $0.57\pm0.03~M_\odot$ \citep{Subasavage2017}, its estimated mass is $M_p = 9.26^{2.64}_{-1.15}~M_{\rm Jup}$ \citep{GC2023AA674A34G}. WD~0141--675~b is also the first planet candidate found around a metal-polluted white dwarf and therefore represents a promising opportunity to investigate the role of outer planets on the metal-pollution of the atmospheres of white dwarfs  \citep{Subasavage2017}.

Lastly, since most stars form in binary star systems, the architecture of planetary orbits in binaries systems will offer valuable information about the influence of stellar companions on both planet formation and subsequent dynamical evolution \citep{Marzari2019}.

\section{Summary}

Astrometry, the oldest method for detecting extrasolar planets, has evolved significantly, achieving unprecedented precision with space-based missions like {\it Gaia} and ground-based interferometry at infrared and radio wavelengths. 
I have revisited the fundamental equations used to model the sky positions of a planet-host star, in the more general case and for {\it Gaia} measurements, and distinguishing between single stars and stars in binary systems. I have also described powerful diagnostic tools, namely the periodogram and the proper motion anomaly, used to investigate the possible presence of planetary companions in two-dimensional astrometric data.

This technique provides critical insights into planetary systems. 
For instance, it can provide all the Keplerian parameters of the planet orbit, as well as the parallax and proper motion of the host star in the case of absolute astrometry. Moreover, the true planet mass can be determined if the host-star mass is known. Most importantly, combining astrometric measurements of the host star and the planetary companion—obtained, for example, through direct detection using infrared interferometers—can enable an accurate determination of the planet’s dynamical mass, which is fundamental not only for constraining the planet's density but also for testing formation scenarios. In the case of planets in binary systems, astrometry provides the three-dimensional configuration of both the binary and planet orbit, enabling a detailed study of their mutual inclination and the system architecture. Such information is crucial for theoretical models, as it can help to understand how the presence of a stellar companion influences the formation and subsequent dynamical evolution of planets.

Astrometry is particularly effective for detecting massive planets ($\sim 1~M_{\rm Jup}$) in long-period orbits ($\gtrsim$ 1 year) around nearby low-mass stars ($M\lesssim0.1~M_{\odot}$), filling gaps left by other techniques. Despite historical challenges, advancements in instrumentation and methodologies have revitalized astrometry, enabling the discovery of exoplanets and the characterization of their orbits. 

The future of this technique is bright, as it will offer unique opportunities to study planetary formation, evolution, and dynamics in binary systems and around white dwarfs. In the optical, Gaia's extended data releases will reveal thousands of Jupiter-mass long-period planets out to distances of $\sim500$~pc.
As astrometric precision continues to improve, future  instruments, such as forthcoming radio arrays, promise to expand the scope of exoplanet discoveries, including low-mass planets and systems that present significant challenges for other exoplanet detection methods.

\clearpage 

\section{Example of data analysis} 

Using public Python libraries, this section exemplifies the process of fitting a planetary orbit to synthetic astrometric data from radio interferometric observations. 
The data provided for the tutorial consists of a series of synthetic positions of the star in equatorial coordinates and the Julian dates corresponding to the middle observing time (Table \ref{tab:data-example}). The barycentric coordinates of the Earth at the observed dates as well as the positional errors are also provided. The simulated data were generated using a model of a planetary companion with $M_p = 3~M_{\rm Jup}$ orbiting around a star with $M_\star = 1~M_\odot$ at a distance of 10 pc. The data were corrupted with astrometric noise assuming positional errors of 50~$\mu$as.

The first step of the analysis consists of searching for periodicities in the data by constructing the circular least-squares periodogram. Then, we  will identify the period of the most prominent peak in the periodogram and define a search window around this period. Using the python library \href{https://lmfit.github.io/lmfit-py/}{{\tt lmfit}}, we will perform a least-squares fitting to find the Keplerian orbit that has the smallest $\chi^2$. 

To start with, we load the data and the python libraries to be used.   

\begin{table*}
\scriptsize 
\caption{Synthetic astrometric data (example.dat).}
\label{tab:data-example} 
\centering 
{
\begin{tabular}{c c c c c c c c c}  
\hline\hline  
Julian   & $X_E$ & $Y_E$ & $Z_E$ & $\alpha(\rm{J}2000.0$)  & $\sigma_\alpha$ & $\delta(\rm{J}2000.0$) & $\sigma_\delta$ \\
Day &  &   &   & [degrees] & [degrees] & [degrees] & [degrees] \\
\hline 
 2458750 & 1.000592574560 &  0.005412924518 & 0.002344110894 & 246.676784671 & 1.4e-08 & -24.440657652 & 1.4e-08 \\
 2458880 & -0.647678460753 &  0.691034620479 & 0.299600493788 & 246.676827892 & 1.4e-08 & -24.440691859 & 1.4e-08 \\
 2459000 & -0.364082478829 &  -0.863474619674 & -0.374253334698 & 246.676790744 & 1.4e-08 & -24.440714328 & 1.4e-08 \\
 2459030 & 0.134854426708 &  -0.917553081433 & -0.397689644839 & 246.676773534 & 1.4e-08 & -24.440718110 & 1.4e-08 \\
 2459060 & 0.599033069363 &  -0.742160041952 & -0.321649568502 & 246.676760010 & 1.4e-08 & -24.440722094 & 1.4e-08 \\
 2459080 & 0.830673529756 &  -0.516970215532 & -0.224027016447 & 246.676754499 & 1.4e-08 & -24.440725238 & 1.4e-08 \\
 2459110 & 0.994931134295 &  -0.078036457797 & -0.033744097168 & 246.676752596 & 1.4e-08 & -24.440730741 & 1.4e-08 \\
 2459130 & 0.961889524966 &  0.233820225628 & 0.101453009978 & 246.676755561 & 1.4e-08 & -24.440735142 & 1.4e-08 \\
 2459160 & 0.703684718032 &  0.640280717549 & 0.277658616932 & 246.676765131 & 1.4e-08 & -24.440742564 & 1.4e-08 \\
 2459190 & 0.256128340946 &  0.876849680197 & 0.380217802486 & 246.676777935 & 1.4e-08 & -24.440750758 & 1.4e-08 \\
 2459250 & -0.711265771417 &  0.638090282456 & 0.276734368496 & 246.676796916 & 1.4e-08 & -24.440767225 & 1.4e-08 \\
 2459330 & -0.829345597337 &  -0.527467017040 & -0.228513865481 & 246.676779227 & 1.4e-08 & -24.440783546 & 1.4e-08 \\
 2459400 & 0.211241308142 &  -0.906606280647 & -0.392854308763 & 246.676739389 & 1.4e-08 & -24.440792854 & 1.4e-08 \\
 2459450 & 0.870142415555 &  -0.455869529512 & -0.197452649556 & 246.676722084 & 1.4e-08 & -24.440800237 & 1.4e-08 \\
 2459490 & 0.979706157783 &  0.150764751424 & 0.065522182551 & 246.676722875 & 1.4e-08 & -24.440808067 & 1.4e-08 \\
 2459538 & 0.528747179774 &  0.764385181251 & 0.331537473568 & 246.676738721 & 1.4e-08 & -24.440820078 & 1.4e-08 \\
 2459570 & -0.000525021799 &  0.905629228388 & 0.392770660055 & 246.676752476 & 1.4e-08 & -24.440829086 & 1.4e-08 \\
 2459610 & -0.645366969566 &  0.692463938653 & 0.300378316305 & 246.676764546 & 1.4e-08 & -24.440840113 & 1.4e-08 \\
 2459630 & -0.869173030793 &  0.449092060014 & 0.194877830598 & 246.676766175 & 1.4e-08 & -24.440845084 & 1.4e-08 \\
 2459730 & -0.376084482924 &  -0.865161763880 & -0.374830789265 & 246.676727740 & 1.4e-08 & -24.440862572 & 1.4e-08 \\
\hline 
\end{tabular}
}
\footnotesize{{\bf Notes.} An  electronic form of this table is available in the online version of the Chapter. }
\end{table*}

\begin{lstlisting}[language=Python]
import numpy as np
import matplotlib.pyplot as plt
from lmfit import Model
import math 
import kepler

# Reading the data 
datafile = 'example.dat'

jd,earth_x,earth_y,earth_z,RA,err_RA,Dec,err_Dec=np.loadtxt(datafile,unpack=True,usecols=(0,1,2,3,4,5,6,7))

# Mean epoch
jd0 = np.mean(jd)

\end{lstlisting}

The measured positions trace the total motion of the star, being the parallax the largest contribution to this motion. We could fit the data with a single-source model that includes proper motion, parallax and reference position and look at the residuals.  Since this single-source model does not include the star motion around the star-planet barycenter, it would result in $\chi^2>>1$. Plotting the residuals of the fit would show that the measured data deviate from the predicted positions by the single-source model, and exhibit a temporal trend consistent with a periodic pattern, indicating the possible presence of a companion. In order to search for such periodicity, we proceed with constructing the circular least-square periodogram.

We first define a function to solve Eq.\ (\ref{eq:Kepler}) and  compute the terms $(\frac{r}{a})\cos \nu$ and $(\frac{r}{a})\sin \nu$ in Eq.\ (\ref{eq:TI_ra}) and (\ref{eq:TI_dec}). Here, we make use of the \href{https://pypi.org/project/kepler.py/}{{\tt kepler}}   library.

\begin{lstlisting}[language=Python]
# Function to compute the time-dependent terms in eqs. (9) and (10)
def KeplerianOrbit(jd,ecc,period,T0):

    M = 2.*np.pi*(jd-T0)/period
    E, cos_nu, sin_nu = kepler.kepler(M, ecc)

    xt = (1.0-ecc*np.cos(E))*cos_nu
    yt = (1.0-ecc*np.cos(E))*sin_nu

    return xt, yt

\end{lstlisting}

Then, we define a function to obtain the total motion of the star, which is given by Eqs.\ (\ref{eq:ra_total2}) and (\ref{eq:dec_total2}) and a function for the single-source model (i.e.\ without a companion). 
These functions call the function {\tt proj\_prlx} (Eqs.\ (\ref{eq:earth_ra}) and (\ref{eq:earth_dec})) which, given a reference position provided in degrees (RA\_0, Dec\_0), returns the correction to the source position due to parallax at the date corresponding to the provided  Earth barycentric coordinates. It should be noted that the correction is calculated for a unit parallax in degrees. 

\begin{lstlisting}[language=Python]
# Single-source model
def single_model(t,t_0,earth,params):

    alpha_0,delta_0,pmx,pmy,prlx = params

    RA_pm  = alpha_0 + pmx*(t-t_0)/365.25
    Dec_pm = delta_0 + pmy*(t-t_0)/365.25

    fa_prlx,fd_prlx = proj_prlx(RA_pm,Dec_pm,earth)

    RA_total  = alpha_0 + pmx*(t - t_0)/365.25 + prlx*fa_prlx
    Dec_total = delta_0 + pmy*(t - t_0)/365.25 + prlx*fd_prlx

    return RA_total, Dec_total
    
# Single-source plus Keplerian model
def Keplerian_model(t,t_0,earth,param):

    alpha_0,delta_0,pmx,pmy,prlx,X1,X2,Y1,Y2,ecc,period,T0 = param

    xt,yt=KeplerianOrbit(t,ecc,period,T0)

    RA_pm  = alpha_0 + pmx*(t-t_0)/365.25
    Dec_pm = delta_0 + pmy*(t-t_0)/365.25

    fa_prlx,fd_prlx = proj_prlx(RA_pm,Dec_pm,earth)

    RA_K  = X1*xt + X2*yt
    Dec_K = Y1*xt + Y2*yt

    RA_total  = alpha_0 + pmx*(t-t_0)/365.25 + prlx*fa_prlx + RA_K
    Dec_total = delta_0 + pmy*(t-t_0)/365.25 + prlx*fd_prlx + Dec_K

    return RA_total, Dec_total, RA_K, Dec_K

# Function to compute the projections of the parallactic ellipse
def proj_prlx(RA_0,Dec_0,earth):
    alpha = np.radians(RA_0)
    delta = np.radians(Dec_0)
    x,y,z = earth
    RA_prlx = (x*np.sin(alpha)-y*np.cos(alpha))/np.cos(delta)                               
    Dec_prlx = x*np.cos(alpha)*np.sin(delta)+y*np.sin(alpha)*np.sin(delta)-z*np.cos(delta)  
    
    return RA_prlx, Dec_prlx
    
\end{lstlisting}

Since the parallax is a common parameter in equations (\ref{eq:ra_total2}) and (\ref{eq:dec_total2}), we need to combine the $\alpha$ and $\delta$ measurements into a single array. This one-dimensional array will have a length equal to twice the number of observed epochs. In addition, we need to create arrays of the same length for the constants and independent variables in Eqs. (\ref{eq:ra_total2}) and (\ref{eq:dec_total2}).

\begin{lstlisting}[language=Python]
# Combining RA and Dec data into a one-dimensional array
RA_Dec = np.append(RA,Dec)
err_RA_Dec = np.append(err_RA,err_Dec)

# Creating arrays for jd0 and jd of the same length 
jd0_comb = np.repeat(jd0,40)
jd_comb = np.append(jd,jd)

# Creating arrays for the Earth barycentric coordinates of the same length 
earth_x_comb = np.append(earth_x,earth_x)
earth_y_comb = np.append(earth_y,earth_y)
earth_z_comb = np.append(earth_z,earth_z)
\end{lstlisting}

To perform the model fitting with {\tt lmfit}, the arrays of the independent and dependent variables should have the same shape. Therefore, we need to define model functions that handle the combined arrays.  

\begin{lstlisting}[language=Python]
# Single-star model for fitting
def single_model_for_fit(X,alpha_0,delta_0,pmx,pmy,prlx):
 
    (t,t_0,RA_Dec,earth_x,earth_y,earth_z)=X

    RA  = RA_Dec[:20]
    Dec = RA_Dec[20:]

    earth = earth_x[:20],earth_y[:20],earth_z[:20]

    fa_prlx,fd_prlx = proj_prlx(RA,Dec,earth)

    RA_total = alpha_0 + pmx*(t[:20]-t_0[:20])/365.25 + prlx*fa_prlx
    Dec_total = delta_0 + pmy*(t[20:]-t_0[20:])/365.25 + prlx*fd_prlx

    return np.append(RA_total, Dec_total)
    
# Single-star plus Keplerian model for fitting
def Keplerian_model_for_fit(X,alpha_0,delta_0,pmx,pmy,prlx,X1,X2,Y1,Y2):

    (t,t_0,RA_Dec,earth_x,earth_y,earth_z,xt_yt)=X

    RA  = RA_Dec[:20]
    Dec = RA_Dec[20:]

    xt = xt_yt[:20]
    yt = xt_yt[20:]

    earth = earth_x[:20],earth_y[:20],earth_z[:20]

    fa_prlx,fd_prlx = proj_prlx(RA,Dec,earth)

    RA_total = alpha_0 + pmx*(t[:20]-t_0[:20])/365.25 + prlx*fa_prlx + X1*xt + X2*yt
    Dec_total = delta_0 + pmy*(t[20:]-t_0[20:])/365.25 + prlx*fd_prlx + Y1*xt + Y2*yt

    return np.append(RA_total, Dec_total)
\end{lstlisting}

We also need to define a function to compute the chi-squared of the fits, $\chi^2$, as well as the reduced chi-squared, $\chi_{\rm red}^2 = \chi^2 / (N-k-1) $, where $N$ is the number of data points and $k$ the number of free parameters in the model. 

\begin{lstlisting}[language=Python]
# Function to compute the Chi for the provided data
def chi_square(time,time_0,RA,err_RA,Dec,err_Dec,earth,param):

    if len(param)==5:
       mod_RA,mod_Dec = single_model(time,time_0,earth,param)
    else:
       mod_RA,mod_Dec,_,_ = Keplerian_model(time,time_0,earth,param)

    chi2_RA = np.sum(((RA-mod_RA)/err_RA)**2)
    chi2_Dec = np.sum(((Dec-mod_Dec)/err_Dec)**2)

    return chi2_RA+chi2_Dec

# Function to compute the reduced Chi for the provided data
def chi2_red(nobs,val,npar):
    nu = 2.*nobs-npar-1.
    red_chi2 = val/nu

    return red_chi2
    
\end{lstlisting}

We load the models and set the model parameters up with initial values.

\begin{lstlisting}[language=Python]   
# Loading Single-source model
amodel = Model(single_model_for_fit)

# Initial values of single-source model paremeters 
aparams = amodel.make_params(alpha_0=240.0,delta_0=-24.0,pmx=-100./1000./3600.,pmy=-100./1000./3600.,prlx=50./1000./3600.)

# Loading Keplerian model
kmodel = Model(Keplerian_model_for_fit)

# Initial values of single-source + Keplerian model paremeters 
kparams = kmodel.make_params(alpha_0=240.0,delta_0=-24.0,pmx=-100./1000./3600.,pmy=-100./1000./3600.,prlx=50./1000./3600.,X1=0.0,X2=0.0,Y1=0.0,Y2=0.0)

\end{lstlisting} 

A least-squares fitting using the single-source model is performed and the reduced chi-square of the null hypothesis, $\chi_0^2$, to be used in Eq.\ (\ref{eq:periodogram}) is obtained. 

\begin{lstlisting}[language=Python]   
# Weights to be used in the fitting 
weights = 1.0/err_RA_Dec

# Earth barycenter coordinates at the observed epochs
earth = earth_x,earth_y,earth_z

# Fitting the single-source model
result = amodel.fit(data=RA_Dec, params=aparams, X=(jd_comb,jd0_comb,RA_Dec,earth_x_comb,earth_y_comb,earth_z_comb),weights=weights)

# Best-fit parameters of the single-source model 
parameters=(result.best_values['alpha_0'],result.best_values['delta_0'],result.best_values['pmx'],result.best_values['pmy'],result.best_values['prlx'])

# Number of observations and number of parameters of the single-source model 
N_obs = len(jd)
k_0 = 5  

# Calculating reduced chi-square of the null hyphothesis 
chi2_0  = chi_square(jd,jd0,RA,err_RA,Dec,err_Dec,earth,parameters)
rchi2_0 = chi2_red(N_obs,chi2_0,k_0)


\end{lstlisting} 

Now, the periodogram power is calculated using Eq.\ (\ref{eq:periodogram}) for each period between 20 and 350 days, in steps of 0.1 days. For this calculation, the eccentricity is fixed to 0. This implies that $\omega$ and $\tau$ are also fixed. Thus, the number of free parameters in the single-source plus Keplerian model is $k_p=8$ (five astrometric parameters plus $a_1$, $\Omega$ and $i$). 

\begin{lstlisting}[language=Python]   
# Period range for the periodogram
periods = np.arange(20,350,0.1)

# Eccentricity and time of periastron passage are fixed
ecc = 0.0
T0 = jd0

# Number of parameters of the single-source plus Keplerian model 
k_p = 8 

# For loop to compute the periodogram power as a function of period (with e, P, tau and omega fixed)
z = np.zeros(len(periods))

for j,p_days in enumerate(periods):

    xt, yt = KeplerianOrbit(jd,ecc,p_days,T0)

    xt_yt = np.append(xt,yt)

    result = kmodel.fit(data=RA_Dec, params=kparams, X=(jd_comb,jd0_comb,RA_Dec,earth_x_comb,earth_y_comb,earth_z_comb,xt_yt),weights=weights)

    parameters=(result.best_values['alpha_0'],result.best_values['delta_0'],result.best_values['pmx'],result.best_values['pmy'],result.best_values['prlx'],result.best_values['X1'],result.best_values['X2'],result.best_values['Y1'],result.best_values['Y2'],ecc,p_days,T0)

    chi2_p = chi_square(jd,jd0,RA,err_RA,Dec,err_Dec,earth,parameters)

    rchi2_p = chi2_red(N_obs,chi2_p,k_p)

    z[j] = ( (rchi2_0 - rchi2_p)/( k_p - k_0) ) / ( rchi2_p/(2.*N_obs - k_p))
    
\end{lstlisting}   

\begin{figure}[!th]
\begin{center}
 \includegraphics[width=0.99\textwidth,angle=0]{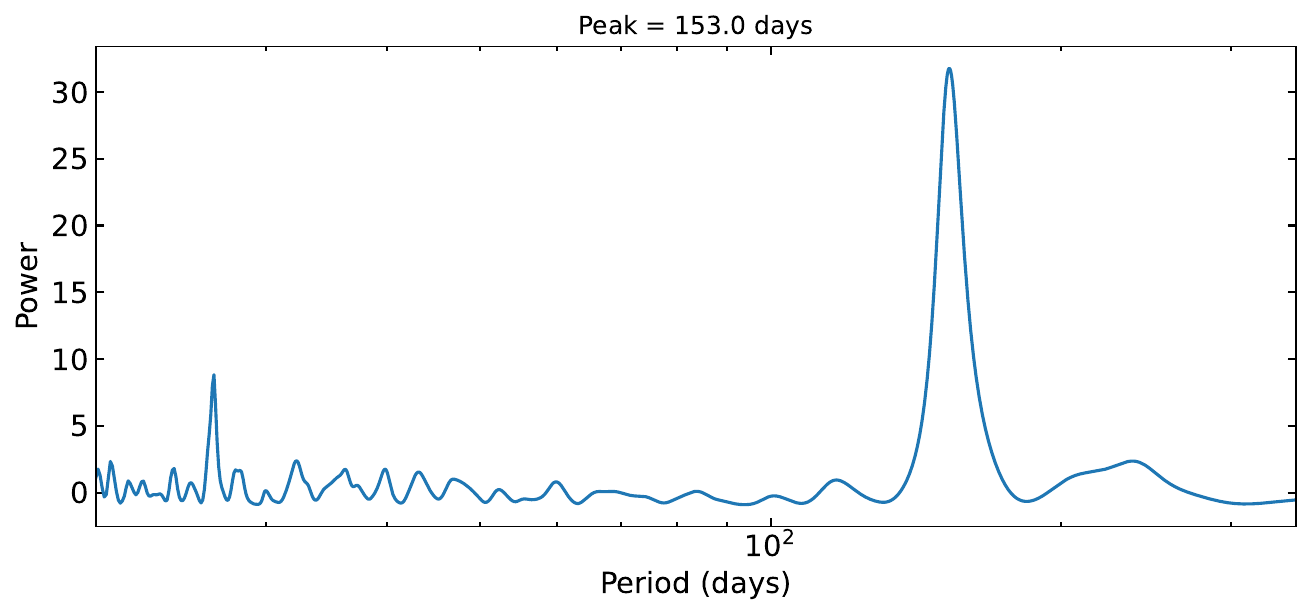}
\caption{Circular least-squares periodogram constructed from the synthetic astrometric data provided in this tutorial. There is a prominent power peak at a period of 153 days. }
\label{fig:example-periodogram}
\end{center}
\end{figure}

We plot the periodogram power as a function of period (see Fig.\ \ref{fig:example-periodogram}). 

\begin{lstlisting}[language=Python]   
# Plotting the periodogram 
plt.close()
fig3 = plt.figure(3,figsize=(10,4))
ax3 = fig3.add_subplot(111)

ax3.plot(periods,z)

# Finding highest peak in the periodogram
i_max  = np.argmax(z)
ax3.set_title(f'Peak = {periods[i_max]:.1f} days')

ax3.set_xscale('log')
ax3.set_xlabel('Period (days)',fontsize=14)
ax3.set_ylabel('Power',fontsize=14)

plt.savefig('periodogram.pdf',bbox_inches='tight')
plt.show()

\end{lstlisting}

Since the periodogram shows a strong peak at 153 days, we specify a search window for the period around this value. We also define search windows for the eccentricity and time of periastron passage. Then, we perform a least-squares fitting for each triplet $(P,e,\tau)$ and compute the $\chi^2$ of the fits. The best-fit solution 
will correspond to the solution that has the smallest $\chi^2$.

\begin{lstlisting}[language=Python]
# Search windows for period, eccentricty and time of periastron passage
periods = np.arange(140,160,1)
eccentricity = np.arange(0,1,0.1)
periastron = np.arange(jd0-300,jd0+300,10)

# Initial value of chi-square
chi2_p_0 = 1.e10

# For loop to compute chi-square for each (P,ecc,T0) triplet 
for p_days in periods:
    for ecc in eccentricity:
        for T0 in periastron:

            xt, yt = KeplerianOrbit(jd,ecc,p_days,T0)

            xt_yt = np.append(xt,yt)

            result = kmodel.fit(data=RA_Dec, params=kparams,\
                                X=(jd_comb,jd0_comb,RA_Dec,\
                                   earth_x_comb,earth_y_comb,\
                                   earth_z_comb,xt_yt),\
                                   weights=weights)

            parameters = (result.best_values['alpha_0'],\
                          result.best_values['delta_0'],\
                          result.best_values['pmx'],\
                          result.best_values['pmy'],\
                          result.best_values['prlx'],\
                          result.best_values['X1'],\
                          result.best_values['X2'],\
                          result.best_values['Y1'],\
                          result.best_values['Y2'],\
                          ecc,p_days,T0)

            chi2_p = chi_square(jd,jd0,RA,err_RA,Dec,err_Dec,earth,parameters)

            if chi2_p < chi2_p_0:

               chi2_p_0 = chi2_p

               b_period = p_days
               b_ecc = ecc
               b_T0 = T0
               b_result = result
               
# Function to convert degrees to milli-arcseconds
def d2m(angle):
    return angle*3600.*1000.
    
# Print best-fit parameters 
print(  f"chi-squared = {chi2_p_0}" )
print(  f"RA_0 = {b_result.best_values['alpha_0']} degrees" )
print(  f"Dec_0 = {b_result.best_values['delta_0']} degrees" )
print(  f"PM RA = {d2m(b_result.best_values['pmx'])} mas/yr" )
print(  f"PM DEC = {d2m(b_result.best_values['pmy'])} mas/yr" )
print(  f"parallax = {d2m(b_result.best_values['prlx'])} mas" )
print(  f"X1  = {b_result.best_values['X1']} degrees" )
print(  f"X2  = {b_result.best_values['X2']} degrees" )
print(  f"Y1  = {b_result.best_values['Y1']} degrees" )
print(  f"Y2  = {b_result.best_values['Y2']} degrees" )
print(  f"period = {b_period} days" )
print(  f"eccentricity = {b_ecc}" )
print(  f"time of periastron passage  = {b_T0}" )

chi-squared = 26.51733394729016
RA_0 = 246.67676762167122 degrees
Dec_0 = -24.44076769164233 degrees
PM RA = -114.0011448883158 mas/yr
PM DEC = -266.95731441316383 mas/yr
parallax = 100.00158403382866 mas
X1  = 2.5560842343669436e-08 degrees
X2  = -1.9661055049830805e-09 degrees
Y1  = -1.987474425404438e-08 degrees
Y2  = 3.5785557814325076e-08 degrees
period = 152 days
eccentricity = 0.30000000000000004
time of periastron passage  = 2459479.4


\end{lstlisting}

Here, we take as the best-fit values of period, eccentricity and time of periastron passage the values corresponding to the point in the 3D grid with the smallest $\chi^2$, while the best-fit values of the astrometric parameters and the Thiele-Innes constants are provided directly by {\tt lmfit}. It should be noted that the grids for $P$, $e$ and $\tau$ that are used in this tutorial are too sparse. In practice, the search should be performed over a denser grid of parameters. Nonetheless, the values that we found are not too far from the true values.   

Now, we plot the best-fit solution of the Keplerian orbit along with the data. To do this, we load Earth barycenter coordinates\footnote{This table is available in electronic form at the online version of the Chapter.} calculated every day for the time range covered by the observations. These coordinates and their corresponding dates are given as input to the single-source plus Keplarian model.  

\begin{lstlisting}[language=Python]
# Barycentric coordinates of the Earth
earthFile = 'earth.dat'
c_jd,c_earth_x,c_earth_y,c_earth_z=np.loadtxt(earthFile,unpack=True,usecols=(0,1,2,3))
c_earth = c_earth_x,c_earth_y,c_earth_z

parameters=(b_result.best_values['alpha_0'],\
            b_result.best_values['delta_0'],\
            b_result.best_values['pmx'],\
            b_result.best_values['pmy'],\
            b_result.best_values['prlx'],\
            b_result.best_values['X1'],\
            b_result.best_values['X2'],\
            b_result.best_values['Y1'],\
            b_result.best_values['Y2'],\
            b_ecc,b_period,b_T0)
            
RA_mc,Dec_mc,RA_mc_K,Dec_mc_K = Keplerian_model(c_jd,jd0,c_earth,parameters)

\end{lstlisting}

In order to plot the Keplerian motion, we remove the fitted parallax and proper motion from the measurement data. 

\begin{lstlisting}[language=Python]
# Removing parallax, proper motion and reference position to the provided data
def removePrlx(t,t_0,alpha,delta,fa_prlx,fd_prlx,params):

    alpha_0,delta_0,pmx,pmy,prlx = params

    d_time = (t-t_0)/365.25

    alpha_wo_prlx = alpha - prlx*fa_prlx - pmx*d_time - alpha_0
    delta_wo_prlx = delta - prlx*fd_prlx - pmy*d_time - delta_0

    return alpha_wo_prlx,delta_wo_prlx

fa_prlx,fd_prlx = proj_prlx(RA,Dec,earth)
            
RA_wo_prlx, Dec_wo_prlx = removePrlx(jd,jd0,RA,Dec,fa_prlx,fd_prlx,parameters[0:5])

\end{lstlisting}

\begin{figure}[!th]
\begin{center}
\includegraphics[width=0.99\textwidth,angle=0]{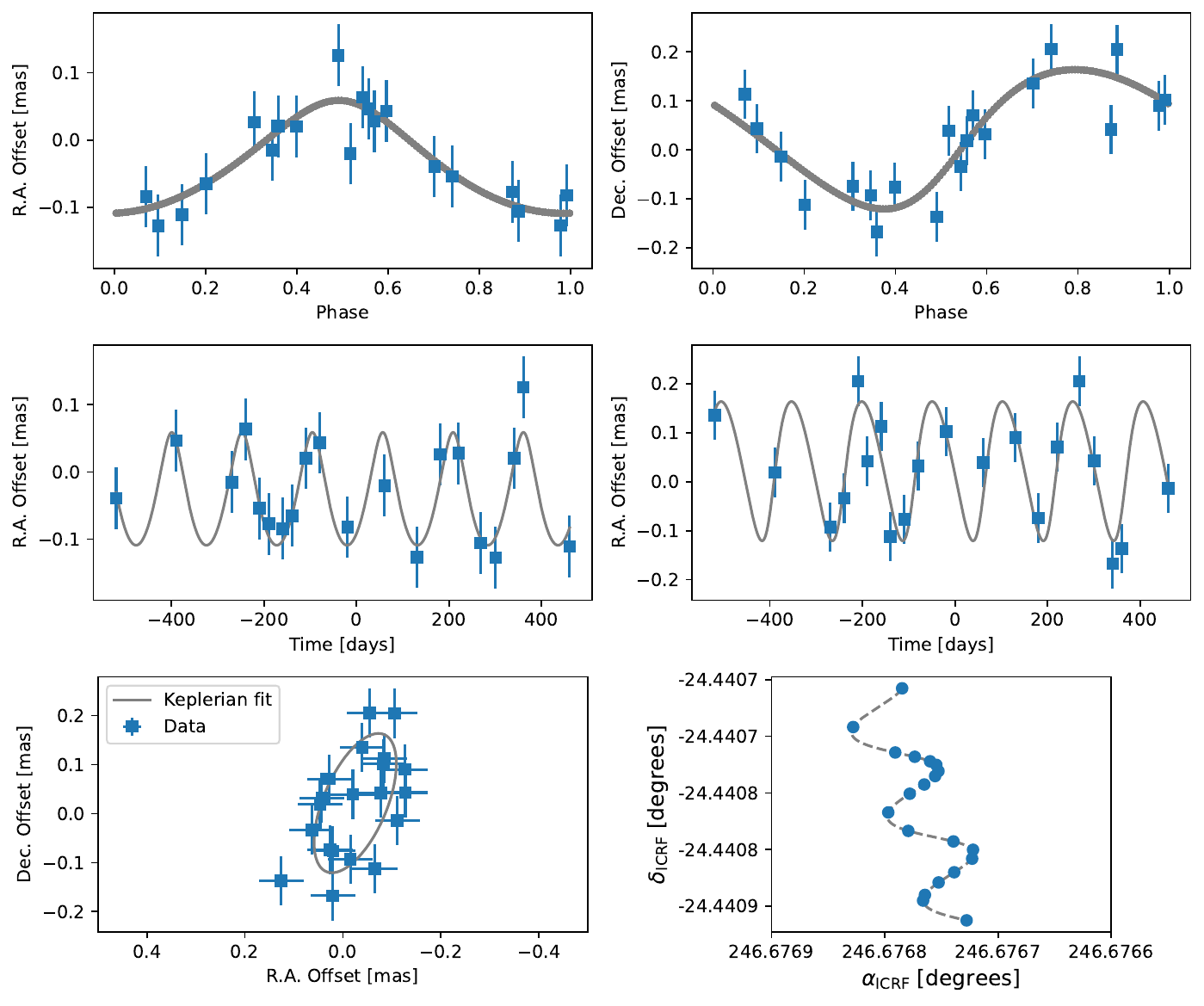}
\caption{Synthetic observed positions of the star after removing the fitted parallax and proper motion. The offsets are relative to the reference position $\alpha$=16$^{\rm h}$26$^{\rm m}$42.4367$^{\rm s}$, $\delta$=--24$^{\rm d}$26$^{\rm m}$26.3255$^{\rm s}$. The top and middle panels show  offsets as a function of the orbital phase and time, respectively. The blue points correspond to the astrometric data, while the solid grey lines show the orbital solution.  The bottom panels show the Keplerian (left) and total (right) motion of the star in the plane of the sky. The dashed gray line shows the single-source plus Keplerian model.  }
\label{fig:fit-Kepler}
\end{center}
\end{figure}

Offsets relative to the reference position are then plotted as a function of the orbital phase. We also plot the orbit the star and the total motion projected on the plane of the sky (see Fig.\ \ref{fig:fit-Kepler}). 

\begin{lstlisting}[language=Python]
# Calculating orbital phase 
phase = ( ( (jd - b_T0 ) / b_period ) + 0.5 ) % 1
c_phase = ( ( (c_jd - b_T0 ) / b_period ) + 0.5 ) % 1

csd = np.cos(np.radians(np.mean(Dec)))

# Plots of the best-fit solution
plt.close()
fig4 = plt.figure(4,figsize=(10, 8))
fig4.subplots_adjust(left=0.1,bottom=0.09,right=0.95,top=0.98,wspace=0.2,hspace=0.3)

ax41 = fig4.add_subplot(321)
ax41.errorbar(phase,d2m(RA_wo_prlx)*csd, yerr=d2m(err_RA)*csd,marker='s',ls='none')
ax41.plot(c_phase,d2m(RA_mc_K)*csd,'.',c='grey',ls='none')
ax41.set_xlabel('Phase')
ax41.set_ylabel('R.A. Offset [mas]')

ax42 = fig4.add_subplot(322)
ax42.errorbar(phase,d2m(Dec_wo_prlx),yerr=d2m(err_Dec),marker='s',ls='none')
ax42.plot(c_phase,d2m(Dec_mc_K),'.',c='grey',ls='none')
ax42.set_xlabel('Phase')
ax42.set_ylabel('Dec. Offset [mas]')

ax43 = fig4.add_subplot(323)
ax43.errorbar(jd-jd0,d2m(RA_wo_prlx)*csd,yerr=d2m(err_RA)*csd,marker='s',ls='none')
ax43.plot(c_jd-jd0,d2m(RA_mc_K)*csd,'-',c='grey')
ax43.set_xlabel('Time [days]')
ax43.set_ylabel('R.A. Offset [mas]')

ax44 = fig4.add_subplot(324)
ax44.errorbar(jd-jd0,d2m(Dec_wo_prlx),yerr=d2m(err_Dec),marker='s',ls='none')
ax44.plot(c_jd-jd0,d2m(Dec_mc_K),'-',c='grey')
ax44.set_xlabel('Time [days]')
ax44.set_ylabel('R.A. Offset [mas]')

ax45 = fig4.add_subplot(325,aspect='equal')
ax45.errorbar(d2m(RA_wo_prlx)*csd,d2m(Dec_wo_prlx),xerr=d2m(err_RA)*csd,yerr=d2m(err_Dec),marker='s',ls='none',label='Data')
ax45.plot(d2m(RA_mc_K)*csd,d2m(Dec_mc_K),'-',color='grey',label='Keplerian fit')
ax45.set_xlim(0.5,-0.5)
ax45.set_xlabel('R.A. Offset [mas]')
ax45.set_ylabel('Dec. Offset [mas]')
ax45.legend(loc=2)

ax46 = fig4.add_subplot(326,aspect='equal')
ax46.errorbar(RA,Dec,xerr=err_RA,yerr=err_Dec,marker='o',ls='none',label='Data')
ax46.plot(RA_mc,Dec_mc,ls='--',c='grey',label='Single-star model')
ax46.set_xlabel(r'$\alpha_{\rm ICRF}$ [degrees]',fontsize=12)
ax46.set_ylabel(r'$\delta_{\rm ICRF}$ [degrees]',fontsize=12)
ax46.set_xlim(246.6769,246.6766)

plt.savefig('plot-Keplerian-fit.pdf',bbox_inches='tight')
plt.show()


\end{lstlisting}

As was mentioned before, the astrometric fitting gives the best-fit values of the Thiele-Innes constants. The Keplerian parameters $a_1$, $\omega$, $\Omega$ and $i$ are obtained from these constants using equations (19.16) to (19.18) of \cite{Green1985}.

\begin{lstlisting}[language=Python]
# Derivation of the angles and semi-major axis from the Thiele-Innes constants
def OrbitParameters(X1,X2,Y1,Y2):

    opo = (math.atan2( (X1-Y2),(X2+Y1) ) )*180./np.pi
    omo = (math.atan2( (X1+Y2),(X2-Y1) ) )*180./np.pi+180.

    i_arg = -1.*( (X1+Y2)*np.sin(np.radians(opo) )/( (X1-Y2)*np.sin(np.radians(omo)) ) )

    omega   = 0.5*( opo  + omo )
    omega_m = 0.5*( opo  - omo ) + 360.

    inclination = (2.*np.arctan(np.sqrt(i_arg)) )*180./np.pi

    a = (X1-Y2)/ ( np.sin(np.radians(opo))*(1.+np.cos(np.radians(inclination)) ) )
    a_mas = d2m(a)

    return a_mas,omega,omega_m,inclination


a1_mas,omega,omegaM,i=OrbitParameters(b_result.best_values['X1'],\
                                      b_result.best_values['X2'],\
                                      b_result.best_values['Y1'],\
                                      b_result.best_values['Y2'])

print(  f"a1 = {a1_mas} mas" )
print(  f"argument of periastron = {omega} degrees" )
print(  f"PA of ascending node = {omegaM} degrees" )
print(  f"inclination = {i} degrees" )

a1 = 0.1584408167839799 mas
argument of periastron = 49.406276142844874 degrees
PA of ascending node = 155.68019909932963 degrees
inclination = 116.8757261121233 degrees
\end{lstlisting}

For comparison, the true parameters of the orbit are $P=150.8$ days, $e=0.23$, $\tau=2459029.075$ days, $a_1=0.15855$ mas, $\omega=56^{\rm o}$, $\Omega=158^{\rm o}$, and $i=110^{\rm o}$. We notice that the semi-major axis of the star orbit (i.e., the astrometric signature) is only a factor of 3 larger than the astrometric errors. Nonetheless, it was still possible to recover the orbit from the astrometric measurements.  

\clearpage

{\bf\Large Table of available instruments}

\begin{table*}[!h]
\caption{Available instruments for astrometric planet searches}
\label{tab:instruments} 
%\centering 
\scriptsize
{
\begin{tabular}{l}  
\hline\hline  
Name    \\
\hline 
Very Long Baseline Array (VLBA) \\
European VLBI Network (EVN) \\ 
\hline
GRAVITY on the Very Large Telescope Interferometer (VLTI)   \\ 
Near-infrared camera (NIRC2) on the Keck II Telescope  \\ 
Wide-field InfraRed Camera (WIRCam) on the 3.6m Canada–France–Hawaii Telescope \\
Center for High Angular Resolution Astronomy (CHARA) Array on Mount Wilson, California \\
Palomar High Angular Resolution Observer (PHARO) camera 5-m  Hale Telescope at the Palomar observatory \\  
\hline
{\it Gaia} Space Observatory \\
Focal Reducer Spectrograph 2 (FORS2) camera  on the 8-m UT1 of the Very Large Telescope (VLT)  \\
High-resolution camera (HRCam) imager on the 4.1-m Southern Astrophysical Research (SOAR) Telescope \\
Gemini Multi-Object Spectrograph (GMOS) on the Gemini-North and -South Telescopes \\ 
Carnegie Astrometric Planet Search Cameras (CAPSCam) on the 2.5-m du Pont Telescope at the Las Campanas Observatory \\ 
\hline 
\end{tabular}
}
\footnotesize{{\bf Notes.} The dividing lines separate instruments operating in the radio, infrared and optical bands, respectively. }
\end{table*}

\bigskip

{\bf\Large Table of available open source tools}

\begin{table*}[!h]
\caption{Available open source tools for astrometric data analysis and orbital fitting}
\label{tab:tools} 
\centering 
\scriptsize
{
\begin{tabular}{l l }  
\hline\hline  
Name   &  \\
\hline 
\href{https://obswww.unige.ch/~delisle/kepmodel/doc/_autosummary/kepmodel.astro.AstroModel.html}{{\tt kepmodel.astro}}  & \url{https://obswww.unige.ch/~delisle/kepmodel/doc/_autosummary/kepmodel.astro.AstroModel.html} \\
\href{https://gitlab.obspm.fr/gaia/nsstools}{{\tt nsstools}} & \url{https://gitlab.obspm.fr/gaia/nsstools} \\
\href{https://orbitize.readthedocs.io/en/latest/}{{\tt orbitize!}}  & \url{https://orbitize.readthedocs.io/en/latest/} \\
\href{https://lmfit.github.io/lmfit-py/}{{\tt lmfit}} & \url{https://lmfit.github.io/lmfit-py/} \\ 
\href{https://pypi.org/project/kepler.py/}{{\tt kepler}}  & \url{https://pypi.org/project/kepler.py/}  \\ 
\href{https://pyastronomy.readthedocs.io/en/latest/modelSuiteDoc/keplerEllipseModelDoc.html}{{\tt KeplerEllipseModel}}  & \url{https://pyastronomy.readthedocs.io/en/latest/modelSuiteDoc/keplerEllipseModelDoc.html} \\ 
\href{https://pyastronomy.readthedocs.io/en/latest/pyaslDoc/aslDoc/keplerOrbit.html}{{\tt Keplerian (two body) orbit}} & \url{https://pyastronomy.readthedocs.io/en/latest/pyaslDoc/aslDoc/keplerOrbit.html} \\ 
\href{https://emcee.readthedocs.io/en/stable/}{{\tt emcee}}  & \url{https://emcee.readthedocs.io/en/stable/} \\ 
\href{http://www.astro.gsu.edu/~gudehus/binary.html}{{\tt Binary Star Combined Solution Package}}  & \url{http://www.astro.gsu.edu/~gudehus/binary.html} \\ 
\hline 
\end{tabular}
}
\end{table*}

\section{Acronyms}

\begin{itemize}

\item Hipparcos --  HIgh Precision PARallax COllecting Satellite

%2D -- Two-dimensional 

%AU -- Astronomical Unit 

\item  ASTRA  -- ASTrometric and phase-Referencing Astronomy 

\item  ARMADA ARrangement for Micro-Arcsecond DifferentialAstrometry

\item CHARA -- Center for High Angular Resolution Astronomy 

\item ESA -- European Space Agency 

\item DR1 -- {\it Gaia} Data Release 1

\item DR3 -- {\it Gaia} Data Release 3

\item LOS -- Line of Sight 

\item LS -- Lomb-Scargle

\item MIRC-X -- Michigan InfraRed Combiner-Exeter 

\item ngVLA -- next-generation Very Large Array 

\item RV -- Radial Velocity 

\item P-type -- Planetary-type 

\item PTI -- Palomar Testbed Interferometer

\item PHASES -- Palomar High-precision Astrometric Search for Exoplanet Systems 

\item PRIMA -- Phase-Referenced Imaging and Microarcsecond Astrometry

\item S-type -- Satellite type 

\item TGAS -- Tycho--Gaia astrometric solution

\item VLBA -- Very Long Baseline Array

\item VLBI -- Very Long Baseline Interferometry

\item VLT -- Very Large Telescope

\item VLTI -- Very Large Telescope Interferometer 

\item WD -- White Dwarfs

\end{itemize}

\section{Acknowledgements}

I thank the anonymous referee for valuable comments that helped to improve this chapter. 
I acknowledge the financial support provided by Secretaría de Ciencia, Humanidades, Tecnología e Innovación (Secihti) through grant CBF-2025-I-201. I also thank Sebastián Terreros-Martínez and Eilitia Juárez-Marín for a detailed reading of an early version of the manuscript and valuable suggestions. Finally, I  thank Salvador Curiel for his helpful discussion on the astrometry method. 

\clearpage
\newpage

\bibliographystyle{mystyle}
\bibliography{biblio.bib}
        
\end{document}